\documentclass[3p]{elsarticle}

\usepackage{algorithm}
\usepackage{algpseudocode}

\usepackage{tabularx}

\usepackage{subfigure}
\usepackage{multirow}

\journal{Journal of Elsevier Computer Networks}

\begin{document}

\begin{frontmatter}

\title{A Quality of Experience-Aware Cross-Layer Architecture for Optimizing Video Streaming Services}

\author{Qahhar Muhammad Qadir$^1$, Alexander A. Kist$^1$ and Zhongwei Zhang$^2$}
\address{$^1$ School of Mechanical and Electrical Engineering\\
safeen.qadir@gmail.com, kist@ieee.org}
\address{$^2$ School of Agricultural, Computational and Environmental Sciences\\
zhongwei.zhang@usq.edu.au}
\address{Faculty of Health, Engineering and Science}
\address{University of Southern Queensland, Australia}



\begin{abstract}
The popularity of the video services on the Internet has evolved various mechanisms that target the Quality of Experience (QoE) optimization of video traffic. The video quality has been enhanced through adapting the sending bitrates. However, rate adaptation alone is not sufficient for maintaining a good video QoE when congestion occurs. This paper presents a \emph{cross-layer architecture} for video streaming that is QoE-aware. It combines adaptation capabilities of video applications and QoE-aware admission control to optimize the trade-off relationship between QoE and the number of admitted sessions. Simulation results showed the efficiency of the proposed architecture in terms of QoE and \emph{number of sessions} compared to two other architectures (\emph{adaptive architecture} and \emph{non-adaptive architecture}).
\end{abstract}


\end{frontmatter}


\section{Introduction}
The increasing popularity of various video services \cite{Cisco2014a} has made studying the Quality of Experience (QoE) important. The ITU-T defines QoE as a measure to evaluate the service quality as perceived by end users \cite{ITU-T2007}. Various technical and non-technical factors affect this new quality measure \cite{Brooks2010}. Among these factors are those related to service preparation, delivery and presentation. This makes the task of maintaining QoE at an acceptable level a challenge. Many solutions have been introduced to tackle the challenge of video traffic quality. However, more promising architectures are required to meet the satisfaction of users and preserve the interest of service providers. This common goal has been targeted by various designs. Different approaches focusing on optimization metrics, scope and adaptation methods are available. They can be deployed individually or jointly to achieve the goal which is called cross-layer design in the later case \cite{Fu2013}. 

Optimization has to resolve the conflict between the interests of end users and network providers. From the end user perspective, maximum quality is expected; whereas low-cost and the number of served users are important from the network providers' perspective. These two can be jointly optimized through an intelligent design. This motivation has promoted the development of cross-layer designs for video transmission that are QoE-aware. The main objective is to utilize network resources efficiently and optimize video quality, throughput or QoE through a joint cooperation between layers and optimization of their parameters. This enables the communication and interaction between layers by allowing one layer to access the data of another layer. For example, having knowledge about the available bandwidth (network layer) helps the sender to adapt the sending rate (application layer). As a result of this cooperation, better quality for as many users as possible can be expected.

Although dynamic rate adaptation enhances video quality, accepting more sessions than a link can accommodate will degrade the quality. We have investigated how rate adaptation of video sources can provide a better QoE in our previous work \cite{Qadir2013}. However, the friendly behavior of the Internet's transport protocol accommodates every video session and makes room for everyone. This causes degradation of QoE of all video sessions in a bottleneck link. Adaptive sources attempt to reduce the transmission rate of all video sources in order to share the available link capacity. This process does not consider how much the QoE at the receiving end will be affected by the adaptation process. To overcome this problem, a mechanism is required to maintain the quality of on-going video sessions.

In this paper, we combine the rate adaptation capability of video applications and our previously proposed QoE-aware admission control \cite{Qadir2015} in a QoE-aware architecture for video streaming. The contribution of this paper is a novel QoE-aware cross-layer architecture for optimizing video streaming services. The proposed architecture addresses the issue of QoE degradation of video traffic in a bottleneck network. In particular, it allows video sources at the application layer to adapt their rate dynamically to the network environment; and the edge of the network at the network layer to protect the quality of active video sessions by controlling the acceptance of new session through a QoE-aware admission control. 

The remainder of the paper is organized as follows: related work is reviewed in Section \ref{Sec:relatedWork}. We introduce our proposed QoE-aware cross-layer architecture in Section \ref{sec:QoE_arachitecture}. The evaluation environment is explained in Section \ref{sec:evaluationEnvironment}. Section \ref{Sec:performanceEvaluation} presents and discusses the results. Finally, Section \ref{Sec:Conclusion} concludes the paper.

\section{Related Work}
\label{Sec:relatedWork}
Extensive research has been done in the area of QoE optimization for video traffic. Some have focused on the optimization of the video's QoE through mechanisms on a single network layer. Classification and survey of these mechanisms can be found in \cite{Qadir2015a,Ernst2014,Maallawi2014}. In this section we only focus on cross-layer designs that have been proposed to optimize the QoE of video traffic.

The optimal rate of competing scalable video sources for QoE optimization has been found in \cite{Goudarzi2012}. Loss-induced distortion caused by video sources has been minimized and QoE has been maximized by obtaining the optimal rate and capturing the exact effect of packet loss in \cite{Goudarzi2010}. The resulting rates from \cite{Goudarzi2012, Goudarzi2010} are proposed to be used by video encoders for online rate adaptation. In \cite{Politis2012}, a rate adaptation scheme and the IEEE 802.21 media independent handover are integrated for a single and scalable coding. In \cite{Khan2006}, the source rate at the application layer and modulation schemes at the radio link layer are optimized for the quality of video streaming using an application-driven objective function. The link adaptation of the high speed downlink packet access and rate adaptation of multimedia applications are integrated in a QoE-based cross-layer framework that is capable of maximizing the QoE \cite{Thakolsri2009}.

Work in \cite{Khalek2012} combines link adaptation based on an online QoS to QoE mapping, buffer-aware source adaptation and video-layer dependent packet retransmission techniques to provide delay-constrained scalable video transmissions with optimized perceptual quality. The impact of power allocation on bit error rate and video source coding structure for Scalable Video Coding (SVC) video over Multi-Input Multi-Output (MIMO) with the aim of QoE maximization has been considered in \cite{Chen2014}.
 
The work presented in \cite{Latre2011} extends the Pre-Congestion Notification (PCN)-based admission control, determines the required redundancy bits for coping with packet loss, and scales video rate to optimize the QoE in multimedia networks. Two different rate adaptation algorithms have been proposed in \cite{Latre2013}; an optimal one to adapt the video rate based on the maximization of service provider's revenue or QoE and a heuristic one based on the utility of each connection. Relying on subjective tests, \cite{Chen2015} proposes a rate adaptation algorithm and devises a threshold-based admission control strategy to maximize the percentage of video users whose QoE constraints can be satisfied. Per user's QoE constraint was defined by the empirical Cumulative Distribution Function (eCDF) of the predicted video quality.

The cross-layer design presented in \cite{Debono2012} has optimized the QoE of the region of interest for mobile physicians by using advanced error concealment techniques. The work in \cite{Singhal2014} has combined the SVC optimization optimum time slicing for layered coded transmission and adaptive Modulation and Coding Scheme (MCS) to trade between the QoE and energy consumption of wireless broadcast receivers.

In \cite{Rugelj2014}, a QoE-aware joint subcarrier algorithm and a power radio algorithm are combined for a QoE-based resource allocation of the heterogeneous Orthogonal Frequency Division Multiple Access (OFDMA) downlink. The model presented in \cite{Ju2012}, efficiently allocates resources for video applications by mapping between Peak-Signal-to-Noise Ratio (PSNR) and Mean Opinion Score (MOS). Admission control and resource reallocation have been deployed in \cite{Ivesic2014} to increase the session admission rate while maintaining an acceptable QoE of multimedia services in Long-Term Evolution (LTE). The authors of \cite{Zhou2013} utilized the QoE prediction model of \cite{Khan2010} to rate the QoE of multimedia services and allocate resources dynamically.

The QoE-aware cross-layer Dynamic Adaptive Streaming over Hypertext Transfer Protocol (HTTP) (DASH) friendly scheduler introduced in \cite{Zhao2014}, allocates wireless resources for each DASH user. The video quality is optimized based on the collected DASH information through an improved SVC to DASH layers mapping and a DASH proxy. The QoE of multi-user adaptive HTTP video in mobile networks has been optimized by adapting the transmission rate of DASH clients that can be supported by lower layers in \cite{ElEssaili2014}. In \cite{Fiedler2009}, an efficient video processing, an advanced realtime scheduling and reduced-reference metrics across the application and network layers are combined as components for a QoE-driven cross-layer design of mobile video systems.

The automatic architecture proposed in \cite{Latre2009} monitors quality related parameters such as packet loss, video frame rate and router queue sizes. Proper actions such as lowering bit rate or adding more Forward Error Correction (FEC) packet are taken to optimize the QoE of multimedia services. In \cite{Oyman2012}, an adaptive cross-layer architecture is presented. The HTTP Adaptive Streaming (HAS)/HTTP-specific media, network QoS and radio QoS are jointly adapted for optimizing the QoE of HAS applications. An end-to-end system for optimizing the QoE in next generation networks is presented in \cite{Zhang2011}. The QoE/QoS parameters at terminals are reported to the QoE management component for analysis and adjustment. The adjusted QoS/QoE of the end user is then sent to the network and source. A joint framework for video transport optimization over the next generation cellular network that overcomes network congestion, cache failure and user mobility issues is designed in \cite{Fu2013}. Path selection, traffic management and frame filtering are mechanisms of the framework for the SVC video streaming over User Datagram Protocol/Real-time Transport Protocol (UDP/RTP). The interface presented in \cite{Mathieu2011} is to enable the ISPs to deliver video contents efficiently and satisfy the user requirement for QoE through dynamic adaptation.

While the discussion has covered similar aspects, \cite{Latre2011,Latre2013,Chen2015} have specifically combined rate adaptation and admission control in a cross-layer design for QoE optimization. In \cite{Latre2011}, the rate of layered video flows is re-scaled and protected through a number of changes to the original PCN. In contrast, our architecture accounts for the QoE of video sessions through a QoE-aware admission control. \cite{Latre2013} integrates an existing standardized Measurement-Based Admission Control (MBAC) system with a novel video rate adaptation, while our work integrates the existing rate adaptation capability of multimedia applications with a QoE-aware admission control. Furthermore, our architecture optimizes the link utilization considering the QoE of video sessions whereas \cite{Latre2013} accounts for QoE as the output of the system. Finally, \cite{Chen2015} incorporates the QoE constraints into the rate adaptation algorithm, but our proposal incorporates QoE in the rate measurement algorithm and admission control.

\section{QoE-Aware Cross-Layer Architecture}
\label{sec:QoE_arachitecture}
Much of the research discussed in the literature proposed rate adaptation for layered videos such as SVC. The video content (base and enhancements layers) generated by a layered encoder is injected to the network, then the network decides whether they are forwarded or dropped. In contrast, this paper proposes online rate adaptation for single layer videos. Instead of sending the whole video content to the network, video sources based on the condition of the network, decide at what rate to transmit the content. By using this strategy, the rate is adjusted on the fly and additional redundant data is not sent to the network during times of congestion. This is in contrast to offline coding which completely relies on coarse network state assumptions \cite{Lie2008}.

Rate adaptation attempts to adapt the sending rate of all video sources to share the available link capacity without considering how much the received QoE will be affected by the adaptation. Therefore, there is a need for a mechanism to control the number of video sessions which can be accommodated with an acceptable QoE. 

\begin{figure}[t]
    \centering
    \includegraphics[width=\textwidth]{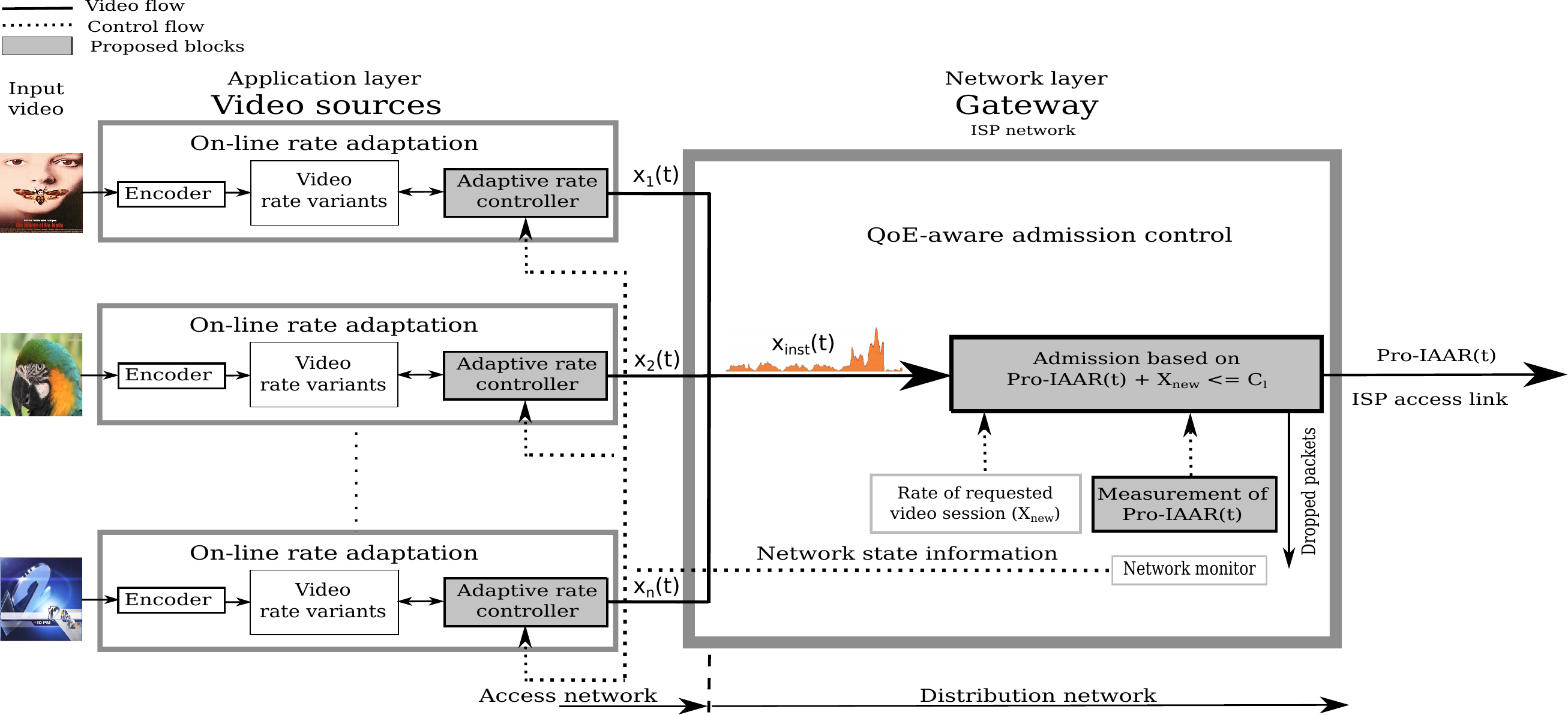}    
    \caption{QoE-aware cross-layer architecture for video traffic}
    \label{fig:Proposed_Framework_paper3_conf_new}
\end{figure}

Figure \ref{fig:Proposed_Framework_paper3_conf_new} shows the proposed architecture which focuses on the optimization of QoE in relation to the \emph{number of sessions} on the ISP access links (ISP links which are directly connected to and controlled by the gateway in Figure \ref{fig:Proposed_Framework_paper3_conf_new}). The video sources share the ISP access links of the distribution network which is controlled by the gateway. The rate adaptation is performed at the application layer and QoE-aware admission control at the network layer. More specifically, The QoE-aware admission control is implemented at the ISP gateway while the sources perform rate adaptation based on the available bandwidth of the ISP access links. 

Unlike current MBACs, the QoE-aware admission control considers the bursty characteristic of video flows as the burstiness of individual video flow can be compensated by the silence of others \cite{Latre2011b}. Bursty traffic refers to inconsistency of the traffic level. It is at high level sometimes while is at low level at some other times. The model and implementation of the QoE-aware measurement algorithm and admission control were presented in \cite{Qadir2015}. The proposed architecture employs parameters from relevant layers; application and network layers in this paper. The key parameters to be considered for the cross-layer optimization are the instantaneous video rate of session \emph{i} at time \emph{t}; \emph{$x_i(t)$} and rate of requested video session; \emph{x}$_{new}$ from the application layer, while at the network layer, the link capacity; \emph{$C_{l}$}, \emph{number of sessions}; \emph{n}, parameter $\beta$ (explained later in this section), and the measured QoE-aware aggregate rate; Proposed-Instantaneous Aggregate Arrival Rate (\emph{Pro-IAAR(t)}) are taken into account. Table \ref{tab:notations} summarizes the notations used for structuring the QoE-aware cross-layer architecture. The architecture assumes that there are efficient and reliable routing protocols to route the video traffic through the ISP intra-domain links once they have been placed on the \emph{ISP access link} by the gateway. It also assumes that there is sufficient bandwidth on the access (between video sources and ISP gateway) and core (Internet) networks.

\begin{table}[t]
\centering
\caption{Notations used for modeling the QoE-aware cross-layer architecture}
\label{tab:notations}
\footnotesize
\begin{tabularx}{\textwidth}{l X}
\hline
Notation 			& Description \\ \hline
 \emph{$x_i(t)$}	& Instantaneous video rate of session \emph{i} at time \emph{t} \\
 \emph{x}$_{new}$	& Rate of requested video session \\
 \emph{n}			& Number of video sessions \\
 \emph{$C_{l}$}		& Link capacity \\
 $\beta$			& A parameter defines the upper limit of the aggregate rate that can exceed \emph{$C_{l}$} while maintaining the QoE of enrolled video sessions \\
 \emph{Pro-IAAR(t)}	& Measured QoE-aware aggregate rate at time \emph{t} \\
 $\mu_s(t)$			& Expected value of the total aggregate rate at time \emph{t} \\
 \emph{p$_{i}(t)$}	& Active or inactive probability of the session \emph{i} at time \emph{t} \\
 $\epsilon$ 		& The positive number defined by the Hoeffding inequality theorem \cite{Hoeffding1963} \\
 $\alpha$ and $\delta$	& Coefficients of $\beta$ model defined by Equation (\ref{eq:beta}) \\
 \emph{k}			& A local variable/counter, where \emph{k} $\in$ \{2-31\} \\
 \emph{QP}			& Quantization parameter of video encoder, where \emph{QP} $\in$ \{2-31\} \\
 
\hline
\end{tabularx}
\end{table}

A user's QoE (in terms of MOS) for video streaming services can be defined by a utility function \cite{Thakolsri2009}. MOS as a function of the aggregate bitrate is given by a simplified utility function in Equation (\ref{eq:QoE_utlityFunction})
\begin{equation}
\label{eq:QoE_utlityFunction}\
U = f(Pro-IAAR(t)), \ f : Pro-IAAR(t) \rightarrow MOS \\
\end{equation}

where \emph{Pro-IAAR(t)} \cite{Qadir2015} is the upper limit of the total aggregate rate that can exceed a specific link capacity considering the QoE of ongoing video sessions and is given by Equation (\ref{eq:pro_iaar})
\begin{equation}
\label{eq:pro_iaar}
Pro\textit{-}IAAR(t) = \mu_s(t)+n\epsilon
\end{equation}

$\mu_s(t)$ is the expected value of the total aggregate rate given by Equation (\ref{eq:expectValue_iaar}) and $\epsilon$ as a positive number of the Hoeffding inequality theorem \cite{Hoeffding1963} is quantified by Equation (\ref{eq:epsilon}) \cite{Qadir2015}. The probability of the session \emph{i} to be active or inactive is represented by \emph{p$_{i}$} in Equation (\ref{eq:expectValue_iaar})

\begin{equation}   
\label{eq:expectValue_iaar}
\mu_{s}(t) = \sum_{i=1}^n x_{i}(t)\  p_{i}(t)
\end{equation}

\begin{equation}   
\label{eq:epsilon}
\ \ \ \ \ \ \ \ \ \ \ \ \ \ \ \ \ \ \ \ \ \ \ \epsilon = \beta \mu_s(t) \frac{n-1}{n} \ \ \ \ \ \ 0<\beta\le1.
\end{equation}

Parameter $\beta$, modeled by Equation (\ref{eq:beta}), defines the upper limit of the total aggregate rate that can exceed \emph{$C_{l}$} while maintaining the QoE of enrolled video sessions. The value of $\beta$ determines the level of video quality. The values of coefficients $\alpha$ and $\delta$ are determined by video contents \cite{Qadir2015}.

\begin{equation}   
\label{eq:beta}
\beta = \alpha + ( \frac{ C_{l} }{\delta * n} ).
\end{equation}

Encoders that provide quality variability such as MPEG-4 can be used to produce different video quality from the video scenes. The rate controller (Figure \ref{fig:Proposed_Framework_paper3_conf_new}) adapts the transmission rate based on \emph{Pro-IAAR(t)}. The load is monitored by the network monitor and \emph{Pro-IAAR(t)} is estimated, the information is then sent back to the rate controller via the acknowledgment packet of Transport Control Protocol (TCP) Friendly Rate Control (TFRC) as an extension of TCP. TFRC can be utilized for this purpose. TFRC is a congestion control mechanism for unicast transmission over the Internet. In addition to fairness when competing with other flows, it has a much lower variation of throughput over time compared with TCP. This makes TFRC more suitable for applications which require smooth sending rate such as video streaming \cite{Floyd2008}. The significance of TFRC for media applications has been growing remarkably \cite{Lie2008}. The rate controller selects a suitable video quality among available bit rates (video rate variants in Figure \ref{fig:Proposed_Framework_paper3_conf_new}) for each Group of Picture (GoP) based on the information on the network state received from the network monitor. An open loop Variable Bit Rate (VBR) controller requires access to the video content and network state information. The Explicit Congestion Notification (ECN) bit in the acknowledgment packet of the TFRC header is utilized for the purpose of network monitoring and thus no additional overhead is introduced. The rate controller at the sender side reduces its transmission rate by selecting a lower video rate variant if ECN 1 is detected in the acknowledgment packet.

The rate controller switches to the next rate by selecting the next quantizer scale at the start of the next GOP. This may delay the new rate up to the duration of one GOP. A leaky bucket can be used to control the target bit rate and allowed bit rate variability. It acts as a virtual buffer, therefore it does not introduce additional delay to video packets. Leaky bucket algorithms are widely used by rate controllers to control traffic to packet-switched and ATM-based networks \cite{Hamdi1997}.

The QoE-aware admission control component measures the network load and based on that makes the admission decision. The new requested session will be admitted only if the sum of \emph{Pro-IAAR(t)} on the link plus \emph{x}$_{new}$ is less than or equal to \emph{$C_{l}$}. The details of a possible scenario is explained in the next paragraph.

\begin{algorithm}[t]
\caption{Implementation of the QoE-aware cross-layer architecture for video admission}
\label{pro-architecture}
\footnotesize  	
\begin{algorithmic}[1]
\Statex $Given\ C_{l},\ n,\ \alpha,\ and\ \delta\ $
\For  {$Every\ video\ session\ request$} 
\State $Compute\ \mu_s(t)\ from\ Equation\ (\ref{eq:expectValue_iaar})$
\State $Compute\ \beta\ from\ Equation\ (\ref{eq:beta})$
\State $Compute\ \epsilon\ from\ Equation\ (\ref{eq:epsilon})$	
\State $Compute\ Pro\textit{-}IAAR(t)\ from\ Equation\ (\ref{eq:pro_iaar})$
\State $k=2$
\State $x_{new}=Highest\ bit\ rate\ (QP=k)$
\If{  $Pro\textit{-}IAAR(t) + x_{new} \le  C_{l} = True$ }
\State $Session\ accepted\ with\ rate\ x_{new}$ 
\State $Send\ the\ QP/k\ that\ satisfies\ accepted\ x_{new},\ to\ the\ source$
\Else
\If{  $k\ \le 31$ } 
\State $Increment\ k$
\State $x_{new}=Next\ bit\ rate\ (QP=k)$
\State $goto\ {\bf line\ 8}$
\Else 
\State $Session\ rejected$
\EndIf 

\EndIf  
\EndFor
\end{algorithmic}
\end{algorithm}

A video source prior to transmitting, sends a request to the ISP gateway indicating its intended sending rate (highest bit rate) as well as other possible bit rates (30 bit rates in total). Existing session signaling protocols such as Session Initiation Protocol (SIP) is currently used by Internet telephone calls and it also can be utilized for video distribution \cite{Rosenberg2002}. The gateway upon receiving the request, calculates $\mu_s(t)$ using Equation (\ref{eq:expectValue_iaar}), $\beta$ using Equation (\ref{eq:beta}), Pro\textit{-}IAAR(t) using Equation (\ref{eq:pro_iaar}) and checks \emph{Pro-IAAR(t)} + \emph{x}$_{new}$ $\le$  \emph{$C_{l}$}. The new session is accepted with its intended bit rate \emph{x$_{new}$} only if the condition meets. If it does not however, the gateway checks the next bit rate (from higher to lower) that satisfies the condition. The gateway acknowledges the potential source should any other bit rate meets the condition which is then adopted by the source. If none of the bit rates satisfies the condition however, the request is rejected. The video sources are able to switch to a higher bit rate after they have been successfully accepted when bandwidth becomes available. Since only the acceptance/rejection admission policy was the target of this paper, post-acceptance bit rate switching was not addressed by our algorithm. The pseudocode for the implementation of the proposed QoE-aware cross-layer architecture for video admission is summarized in Algorithm \ref{pro-architecture}.

QoE is included into Algorithm \ref{pro-architecture} through parameter $\beta$ which controls the total bitrate on a specific link based on the QoE of current sessions. On the other hand, the rate controller makes the architecture flexible by offering 30 different bit rates-with preference from high to low-assuming that they do not cause noticeable artifacts.

Algorithm \ref{pro-architecture} is jointly implemented by the video sources and ISP gateway relying on the available communication messages of the TCP/IP protocol suite for showing the interest to send, notification of the sender and network monitoring as explained earlier in this section. It therefore, does not demand additional requirements. We assume that each media content is encoded with 30 video rate variants. This allows for a wide range of playback rates (30) exploiting the capability of the ffmpeg encoder. The assumption is justifiable for video streaming services and the dropping cost of storage on media servers. Other studies have chosen videos files dynamically in response to channel conditions and screen forms under a limited storage budget through intelligent algorithms \cite{Zhang2013b}.

Using Big O notation metric, the complexity of Algorithm \ref{pro-architecture} is determined by counter $k$ of the iteration loop in line 12 as well as fundamental operations in lines 2, 3, 4, 5, 6, 7, 8, 9, 10 and 17. This describes the worst-case scenario when the condition in line 8 is not satisfied. The time complexity of our algorithm is linear to the counter $k$, i.e.

\begin{equation}   
\label{eq:complexity}
T(k) = 10+1+(k+1)+3k 
\end{equation} 

\begin{equation}   
\label{eq:complexity_final}
T(k) = 12+4k
\end{equation}

that is to say, $T$($k$) $\sim$ ~O($k$). The space complexity of the algorithm such as memory requirement, is insignificant due to the large storage capacity of modern routers.

Each of the on-line rate adaptation and QoE-aware admission control was implemented and investigated separately in \cite{Qadir2013} and \cite{Qadir2015} respectively. In this paper, the functionalities of both components are combined and evaluated within our architecture.

\section{Evaluation Environment}
\label{sec:evaluationEnvironment}
This section describes the settings of the evaluation environment for testing the performance of our architecture. Two video clips with different resolutions were used. The objective of having different video resolutions was to see the impact of video frame size on the performance metrics not to compare these two resolutions. The description of the video contents as well as coding and network parameters are shown in Tables \ref{tab:rateAdaptation_video} and \ref{tab:parameter_settings_rateAdaptation} respectively.

\begin{table}[t]
\centering
\caption{Description of the video sequences used in this paper}
\label{tab:rateAdaptation_video}
\footnotesize
\begin{tabular}{ l l l }
\hline
 Description & Video sequence 1 & Video sequence 2 \\ \hline
 Name & Mother And Daughter & Grandma \\
      & (MAD) & \\
 Description & A mother and daughter & A woman speaking at \\
             & speaking at low motion & low motion \\
 Frame size & CIF (352$\times$288) & QCIF (176$\times$144) \\
 Duration(second) & 30 & 28  \\ 
 Number of frames & 900 & 870 \\
  \hline
\end{tabular}
\end{table}

\begin{table}[t]
\centering
\caption{Simulation parameters}
\label{tab:parameter_settings_rateAdaptation}
\footnotesize
\begin{tabular}{ l l l }
\hline
 & Parameter & Value \\ \hline
\multirow{5}{*}{Encoder} & Frame rate(fps) & 30 \\
 & GoP & 30 \\ 
 & Video quantizer scale & 2 (\emph{non-adaptive architecture} traffic) \\
 & & 2-31 (\emph{adaptive architecture} \& \\ 
 & & \emph{cross-layer architecture} traffic) \\ \hline
\multirow{15}{*}{Network} & \emph{$C_{l}$}(Mbps) & 32 (MAD) \\
 & & 7 (Grandma) \\ 
 & $\beta$ & 0.9 (MAD, \emph{cross-layer architecture}) \\
 & & 0.78 (Grandma, \emph{cross-layer architecture}) \\  
 & VBR sources & 24 \\
 & FTP sources & 48 \\
 & Packet size(byte) & 1052 \\
 & UDP header size(byte) & 8 \\
 & IP header size(byte) & 20 \\
 & Queue size(packet) & 300 (MAD) \\
 & & 100 (Grandma) \\
 & Link delay(millisecond) & 1 \\
 & Queue management & Droptail \\ 
 & Queue discipline & FIFO (First In First Out) \\ 
 & Simulation time(second) & 500 \\ 
\hline
\end{tabular}
\end{table}

\begin{figure}[t]
\centering
\includegraphics{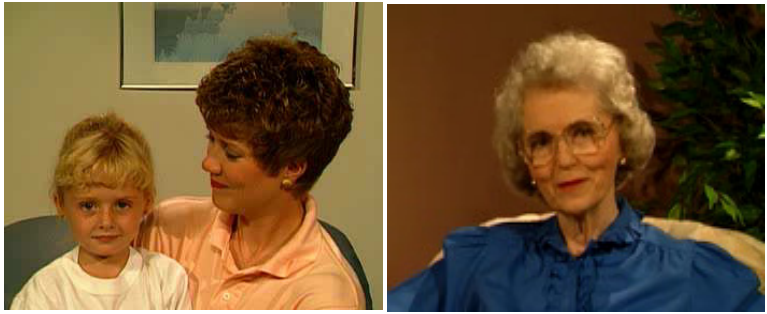}
\caption{Snapshots of the video sequences used in this paper, MAD (left) and Grandma (right)}
\label{fig:MAD_Grandma.eps}
\end{figure}

NS-2 \cite{ns2} was used to simulate the 30 second Common Intermediate Format (CIF) Mother And Daughter (MAD) and 28 second Quarter Common Intermediate Format (QCIF) Grandma video sequences shown in Figure \ref{fig:MAD_Grandma.eps}. In this paper, QCIF (176$\times$144) and CIF (352$\times$288) are specifically chosen as acceptable video formats for most video capable devices such as handsets, mobiles and videoconferencing systems delivered on telephone lines \cite{Khan2012,Qadir2015}. Whereas current mobile devices support bigger sizes, QCIF and CIF make packet level simulation practical. Other resolutions can be applied to the proposed architecture with different values of coefficients $\alpha$ and $\delta$ \cite{Qadir2015}.

\begin{figure}[t]
    \centering
    \includegraphics[width=80mm,height=30mm]{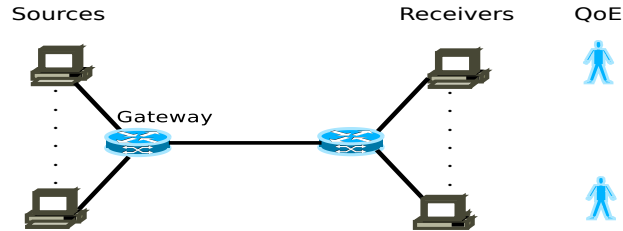}
    \caption{Topology scenario considered in this paper}
    \label{fig:ScenarioDiagram_computerNetworks}
\end{figure}

The topology shown in Figure \ref{fig:ScenarioDiagram_computerNetworks} with a bottleneck link was considered for evaluating the performance of the architecture. A maximum of (24) video sources were competing for the capacity of the link. As the video sources were always active in this paper, \emph{p$_{i}$} was set to 1. There were also (48) File Transfer Protocol (FTP) sources active. The FTP sessions created background traffic and video sessions started randomly during the first 20-50 second of the simulation. The objective was to have a more realistic scenario where other traffic exists in the same network along with the video traffic. In total 500 seconds were simulated. 

The proposed QoE-aware architecture (referred to as \emph{cross-layer architecture}) was compared in details to an architecture (referred to as \emph{adaptive architecture}) in which video sources adapt their bit rates only, while in the \emph{cross-layer architecture}, the gateway implements the QoE-aware admission control in addition to the rate adaptation by video sources. Both architectures were then compared to a \emph{non-adaptive architecture} in which the video flow is sent without rate adaptation and QoE-aware admission control. Similar simulation parameters and environment were used for the comparison. 

Evalvid-RA \cite{Lie2008} was used to implement on-line rate adaptation from different encoded videos each with a valid range (2-31) of Quantization Parameter (QP). A lower QP generates a higher bit rate and better video quality. The MAD and Grandma video sequences were utilized by the NS-2 simulator through video trace files using EvalVid-RA. The \emph{non-adaptive} videos were encoded with QP of 2 whereas the \emph{cross-layer architecture} and \emph{adaptive} videos with QP between 2-31 using ffmpeg encoder \cite{ffmpeg2004} (thirty video sequences with different bit rates).

The video sessions were competing for the \emph{$C_{l}$} described in Table \ref{tab:parameter_settings_rateAdaptation}. The \emph{cross-layer architecture} was configured so that new session was requested randomly within every second of the simulation time and accepted if there is enough bandwidth, i.e the condition \emph{Pro-IAAR(t)} + \emph{x}$_{new}$ $\le$  \emph{$C_{l}$} is satisfied. The arrival of new sessions in this manner avoids the possibility of having ``flash crowd" phenomenon when numerous sessions arrive at the same time \cite{Chen2014a}. Whereas in the \emph{adaptive architecture} and \emph{non-adaptive architecture}, all sessions were admitted for each simulation run, this paper considered only video sessions that were successfully decoded and played back by the receiver (through ffmpeg decoder) as the metric \emph{number of sessions}. Both alternative architectures allow for more sessions, but only those which are decoded and played back successfully by the receiver were taken into account. For simplicity, the maximum number of competing video sessions was limited to 24 sessions. 

MOS, \emph{n}, packet loss ratio and delay were measured as performance metrics. There are no significant jitter requirements for streaming video (the target traffic of this paper) \cite{Szigeti2004}. The studied metrics for both resolutions are plotted next to each other for the sake of convenience not comparison. Cumulative Distribution Functions (CDF) of the means were calculated for the video flows for each metric over 30 runs. MOS was measured using Evalvid \cite{Gross2004}. Evalvid provides a set of tools to analyze and evaluate video quality by means of PSNR and MOS metrics. The Evalvid MOS metric used in this paper calculates the average MOS value of all frames for the entire video with a number between 1 and 5. This tool has widely been used for the similar purpose \cite{Papadimitriou2007,Li2010,Zheng2015,Khan2010a}. Parameter $\beta$ was experimentally found to be 0.9 for the MAD video sequence and 0.78 for the Grandma video sequence. It was also calculated using Equation (\ref{eq:beta}). The values of coefficients ($\alpha$ and $\delta$) were adopted from \cite{Qadir2015} (Table IV for MAD sequence and last paragraph of Section VIII for Grandma sequence). Experimental and calculated $\beta$ are illustrated in Table \ref{tab:beta_calculation}.

\begin{table}[b]
\centering
\caption{Calculation of $\beta$}
\label{tab:beta_calculation}
\footnotesize
\begin{tabular}{ l l l l l l l }
\hline
Video sequence & $\beta$ (Experimental) & $\beta$ (Equation \ref{eq:beta}) & $\alpha$ & $\delta$ & \emph{$C_{l}$}(Mbps) & mean \emph{n} \\ \hline
MAD (CIF) & 0.9 & 0.84 & -0.54 & 0.96 & 32 & 24 \\
Grandma (QCIF) & 0.78 & 0.775 & -0.1 & 0.4 & 7 & 20 \\
\hline
\end{tabular}
\end{table}

\section{Performance Evaluation}
\label{Sec:performanceEvaluation}
In this section, the performance of the video flows in the \emph{cross-layer architecture} is compared to the video flows in the \emph{adaptive architecture} in terms of MOS, number of successfully decoded sessions and delay. Finally, a comparison between the video flows in the \emph{non-adaptive architecture}, \emph{adaptive architecture} and \emph{cross-layer architecture} is made.

\begin{figure}[t]
\centering
\subfigure[MAD (CIF)]{%
\includegraphics[height=4cm,width=5.7cm]{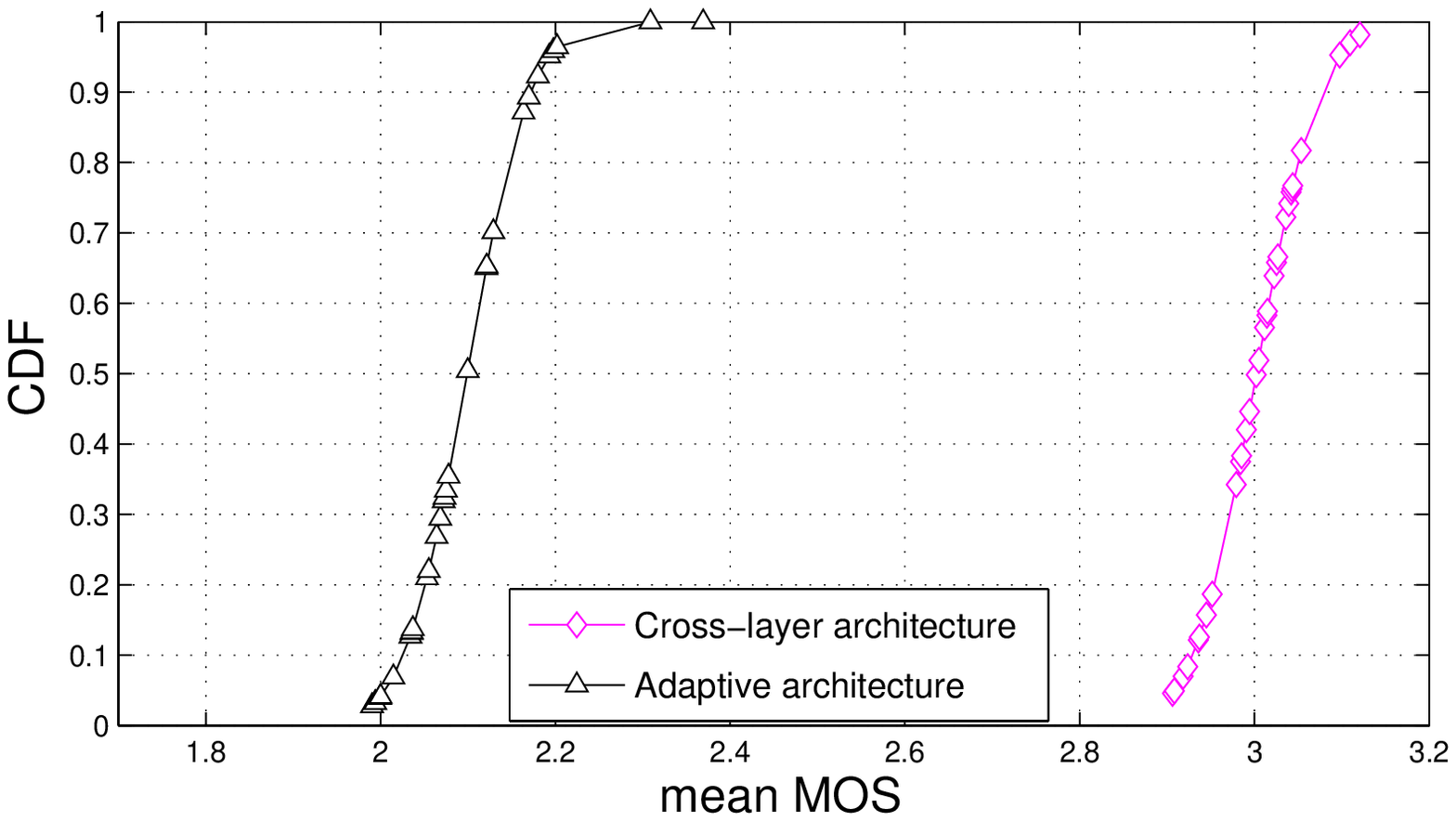}
\label{fig:meanMOS_cross_adaptive_MAD}}
\quad
\subfigure[Grandma (QCIF)]{%
\includegraphics[height=4cm,width=5.7cm]{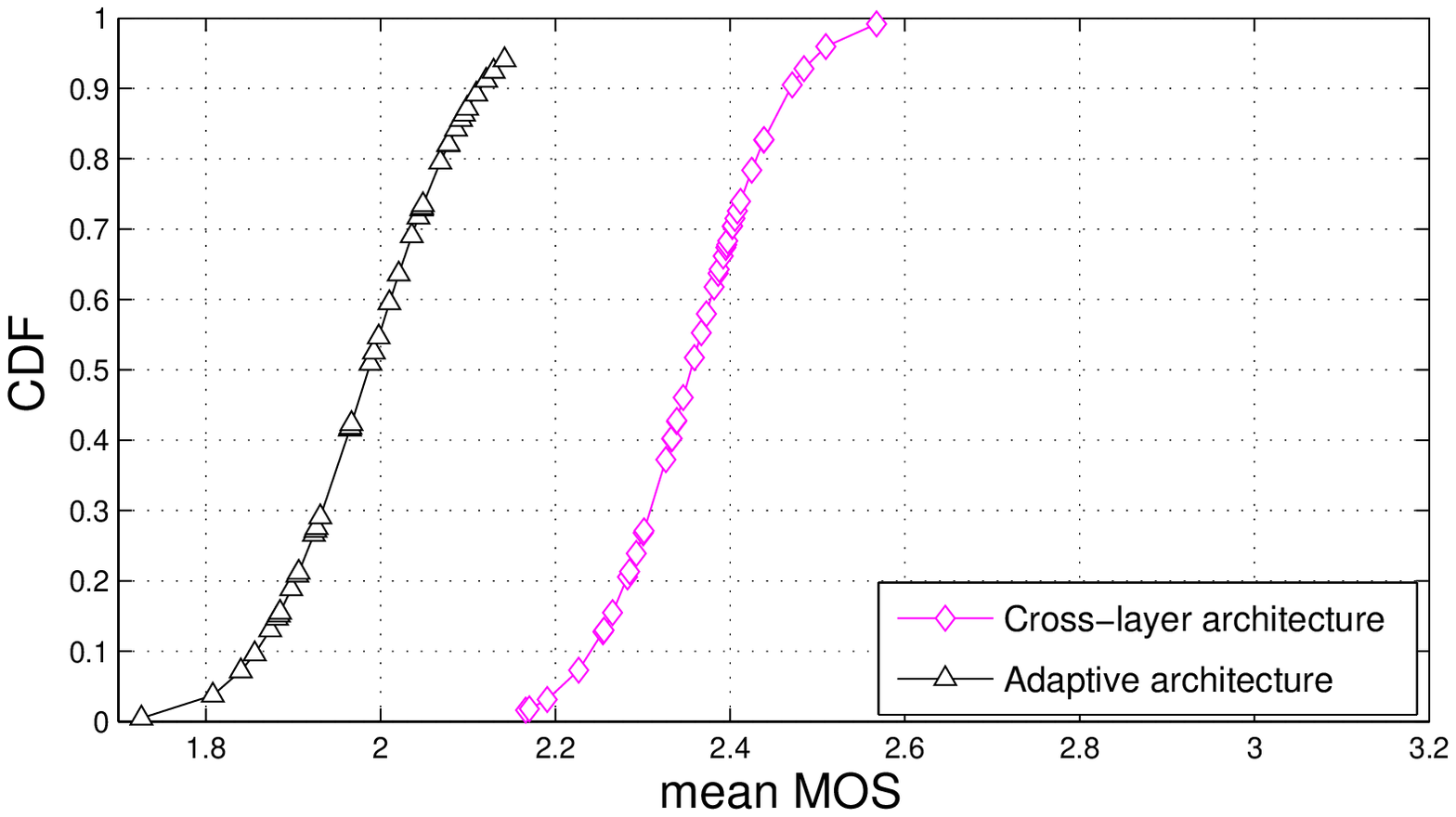}
\label{fig:meanMOS_cross_adaptive_Grandma}}
\caption{CDF of the mean MOS of the video flows in the \emph{cross-layer architecture} and \emph{adaptive architecture} for MAD and Grandma sequences}
\label{fig:meanMOS_cross_adaptive}
\end{figure}

The CDF of the mean MOS of the video flows in the \emph{cross-layer architecture} and \emph{adaptive architecture} for both resolutions are plotted in Figure \ref{fig:meanMOS_cross_adaptive}. MOS enhancement of the video flows delivered through the proposed \emph{cross-layer architecture} can be seen for both resolutions. The difference between the graphs shows that the result depends on the resolution. The mean MOS of the video flows in the \emph{adaptive architecture} was enhanced by the \emph{cross-layer architecture} from 1.98 to 2.35 for the QCIF resolution and from 2.09 to 3 for the CIF resolution. Although, the enhancement of the QCIF resolution can be considered minor, it is substantial for the CIF resolution as the MOS changes from bad to fair according to the absolute mapping in \cite{Ohm2004} \cite{Stankiewicz2011}. As the maximum possible MOS for any multimedia service in practice is 4.5 \cite{Thakolsri2009}, even the slight enhancement of the QCIF MOS by the \emph{cross-layer architecture} can make a difference.

It is worthwhile to mention that the performance of the QoE-aware rate measurement algorithm and associated admission control were more pronounced in terms of MOS when they were evaluated among video flows only in \cite{Qadir2015}. In this paper, FTP traffic is included as a background traffic. Rate adaptation implemented by the video sources lets the video flows respond to the FTP flows by adapting their sending rates. This resulted in a lower MOS compared to the MOS of the video flows in \cite{Qadir2015} in which FTP flows were not considered.

\begin{figure}[t]
\centering
\subfigure[MAD (CIF)]{%
\includegraphics[height=4cm,width=5.7cm]{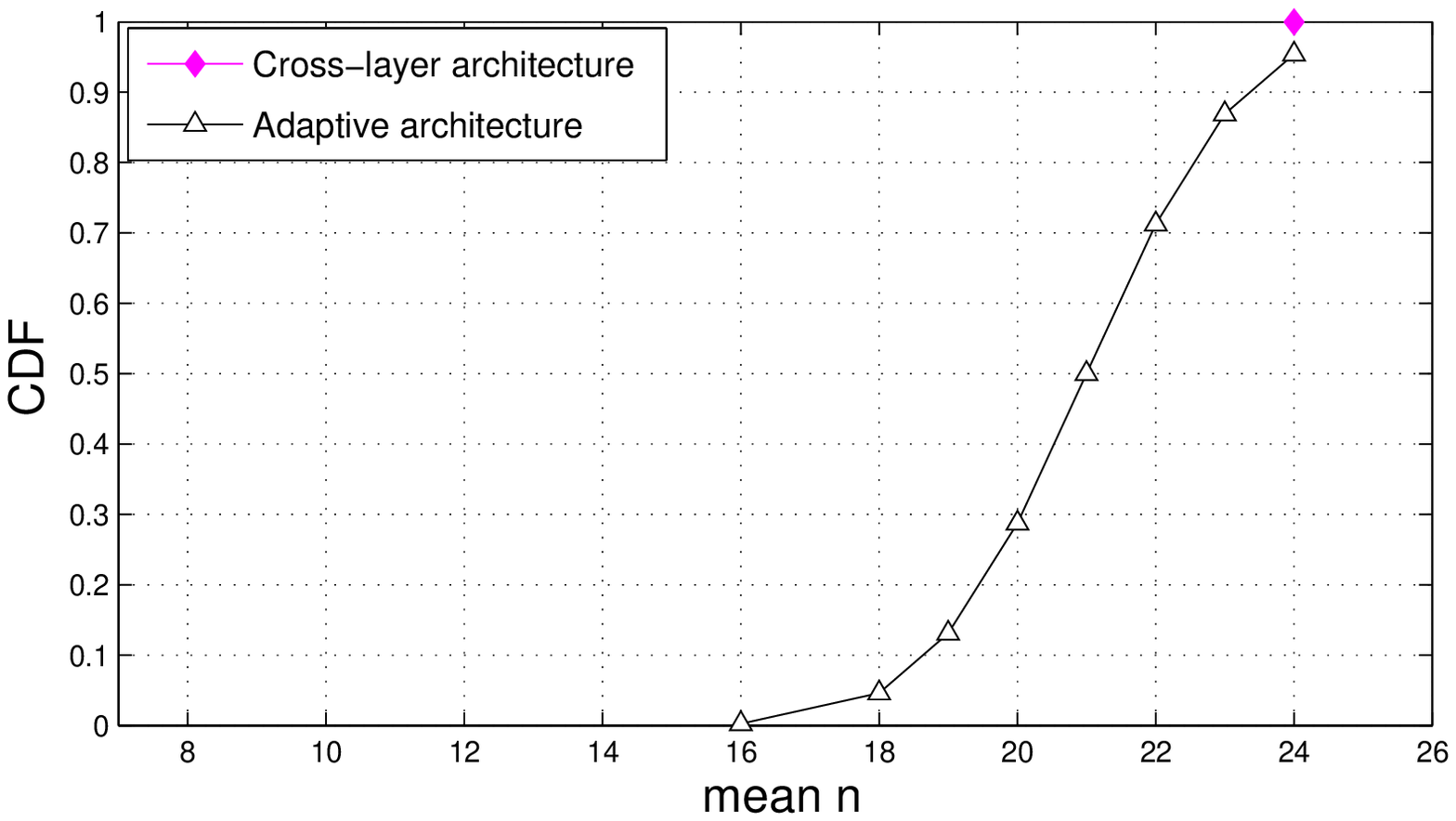}
\label{fig:meanSession_cross_adap_MAD}}
\quad
\subfigure[Grandma (QCIF)]{%
\includegraphics[height=4cm,width=5.7cm]{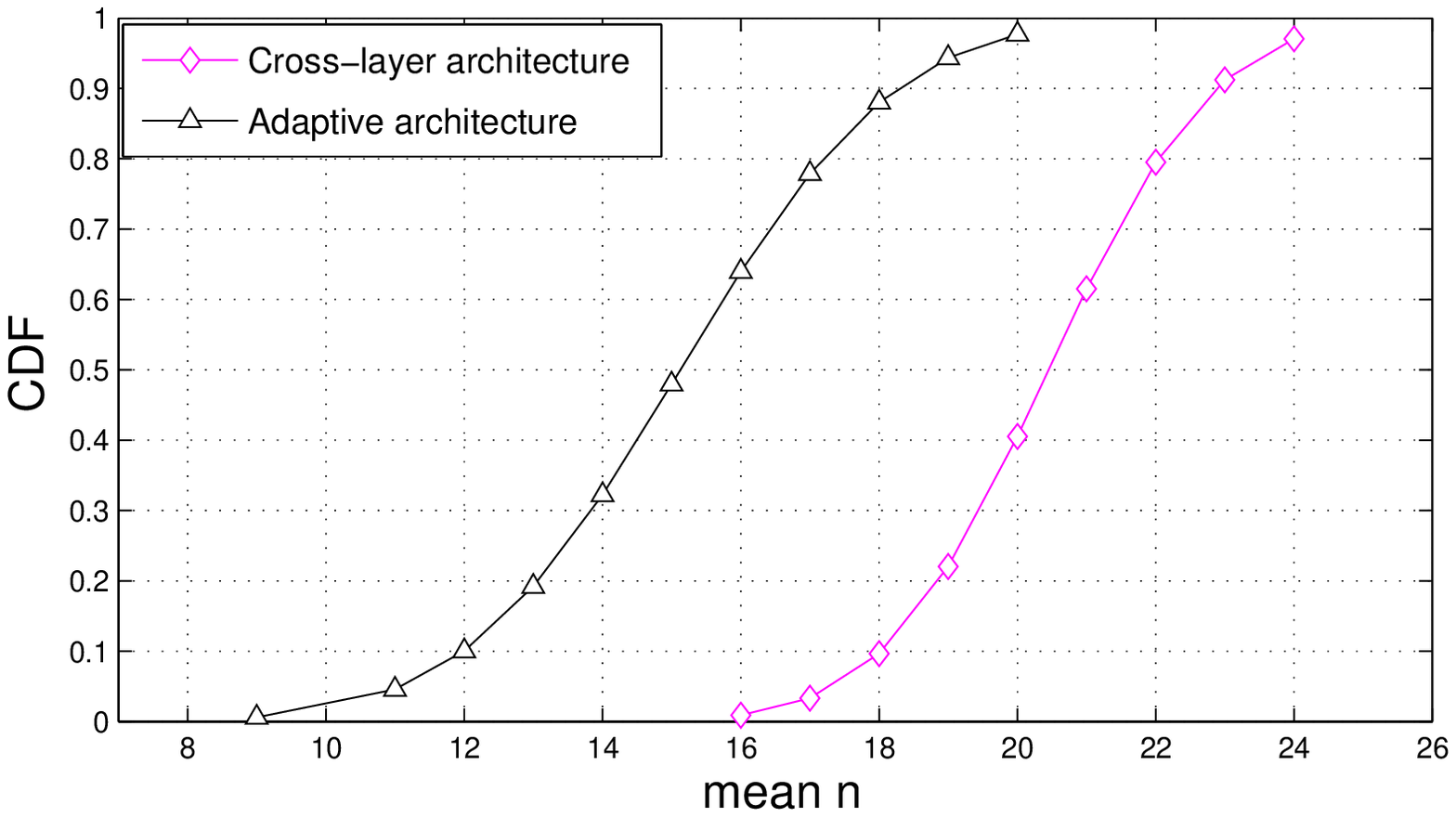}
\label{fig:meanSession_cross_adap_Grandma}}
\caption{CDF of the mean \emph{number of sessions} in the \emph{cross-layer architecture} and \emph{adaptive architecture} for MAD and Grandma sequences}
\label{fig:meanSession_cross_adap}
\end{figure}

\begin{figure}[t]
\centering
\subfigure[MAD (CIF)]{%
\includegraphics[height=4cm,width=5.7cm]{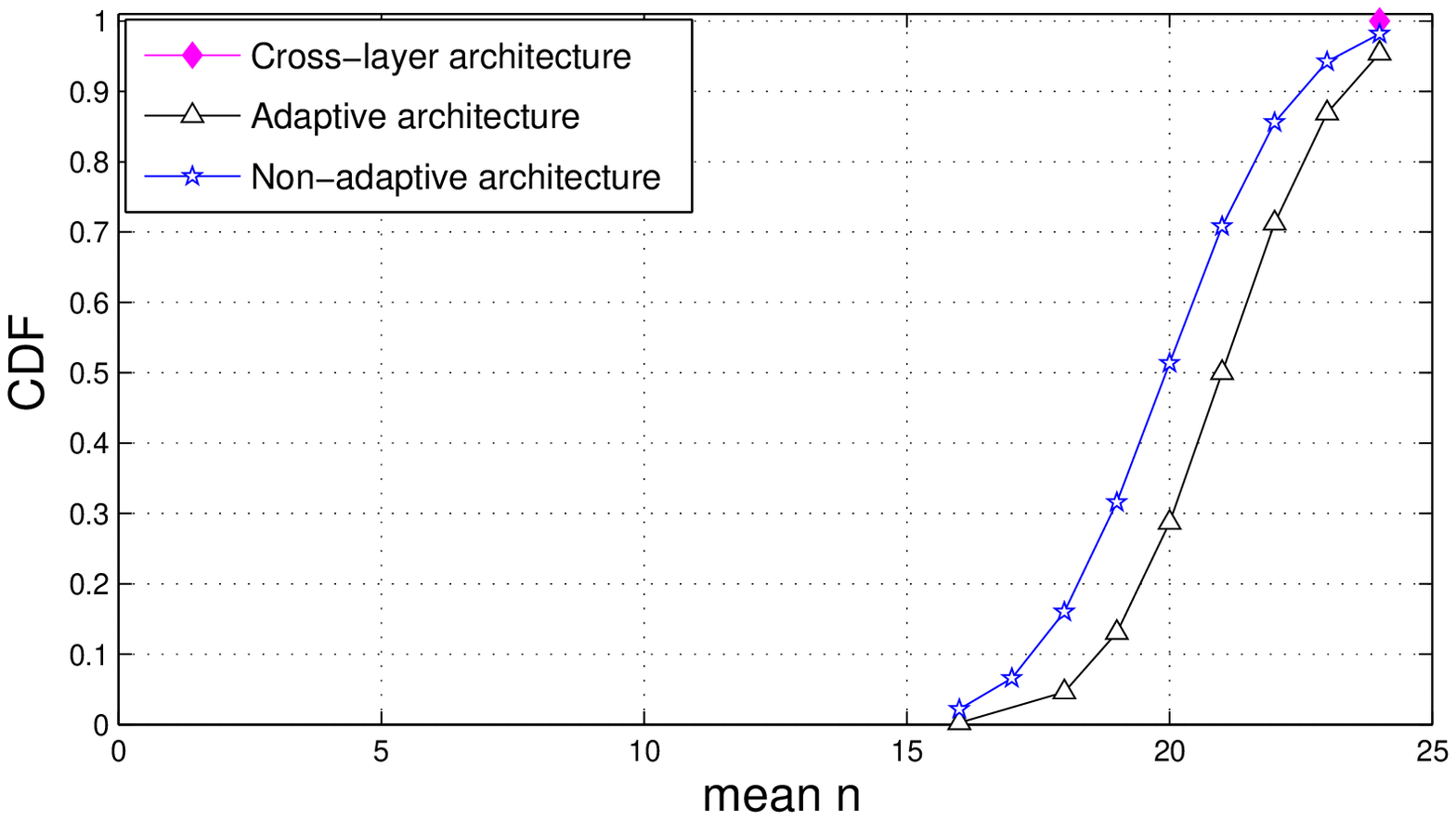}
\label{fig:meanSession_all_adap_MAD}}
\quad
\subfigure[Grandma (QCIF)]{%
\includegraphics[height=4cm,width=5.7cm]{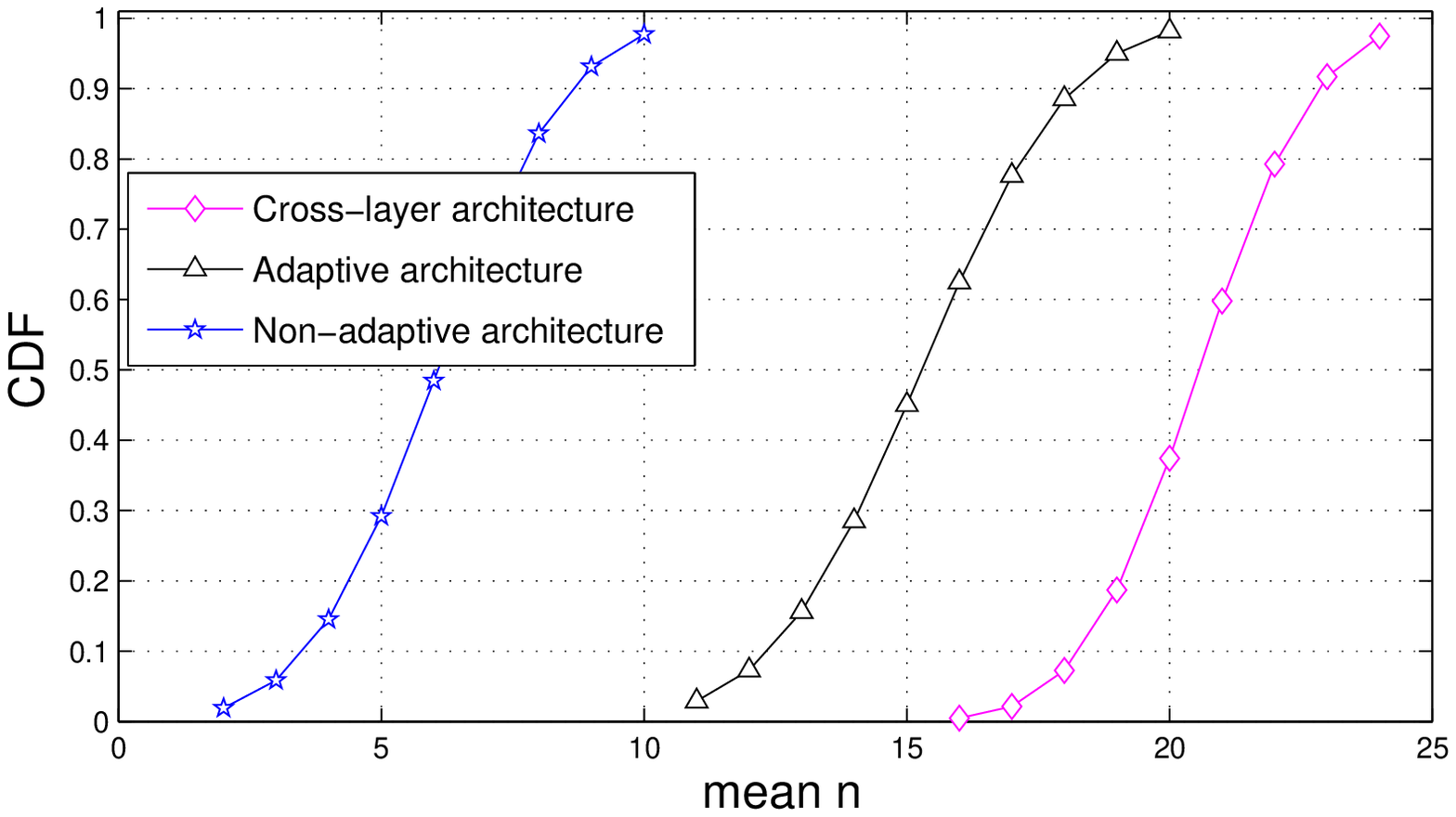}
\label{fig:meanSession_all_Grandma}}
\caption{CDF of the mean \emph{number of sessions} in the \emph{cross-layer architecture}, \emph{adaptive architecture} and \emph{non-adaptive architecture} for MAD and Grandma sequences}
\label{fig:meanSession_all}
\end{figure}

As the main target of this paper is to optimize the QoE-\emph{number of sessions} trade-off, we can not consider the MOS of the video sessions alone. To account for this, the number of successfully decoded video sessions was measured for the \emph{cross-layer architecture}, \emph{adaptive architecture} and \emph{non-adaptive architecture}. This is plotted for both resolutions in Figure \ref{fig:meanSession_cross_adap} and Figure \ref{fig:meanSession_all}. Although, all 24 video flows were active in the \emph{adaptive architecture} and \emph{non-adaptive architecture}, an average of 15 QCIF/21 CIF sessions and 5.9 QCIF/19.9 CIF sessions were successfully decoded by the receivers respectively. This is due to the fact that being adaptive, the video sources in the \emph{adaptive architecture} send data in cooperative manners. Thus not all the video frames were sent into the network due to insufficient bandwidth and availability of other traffic (FTP) in the network. In contrast, an average of 20 QCIF and all 24 CIF videos sessions were successfully decoded when delivered on the \emph{cross-layer architecture}. Although the FTP flows again existed, the video sessions were better managed by the QoE-aware admission control and thus more sessions were accommodated.

It can also be noticed in Figure \ref{fig:meanSession_cross_adap} that the \emph{number of sessions} in the \emph{cross-layer architecture} is not resolution dependent as 5 more QCIF and 3 more CIF sessions are accommodated. As stated in Section \ref{sec:evaluationEnvironment}, due to each resolution's specific simulation settings, the mean MOS and mean \emph{number of sessions} of the two resolutions were not compared to each other.

\begin{figure}[t]
\centering
\subfigure[MAD (CIF)]{%
\includegraphics[height=4cm,width=5.7cm]{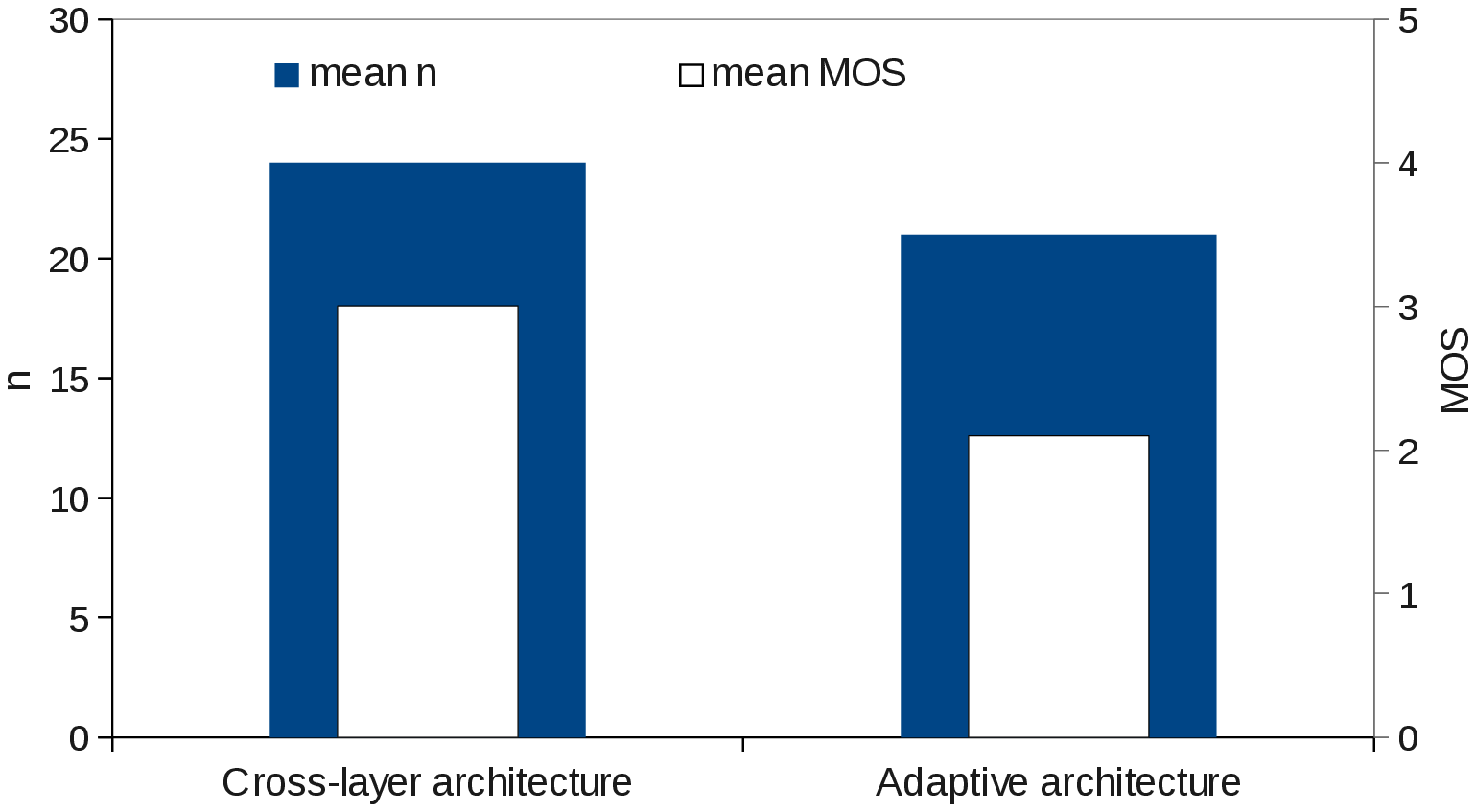}
\label{fig:mos-Session_crossANDadap_MAD}}
\quad
\subfigure[Grandma (QCIF)]{%
\includegraphics[height=4cm,width=5.7cm]{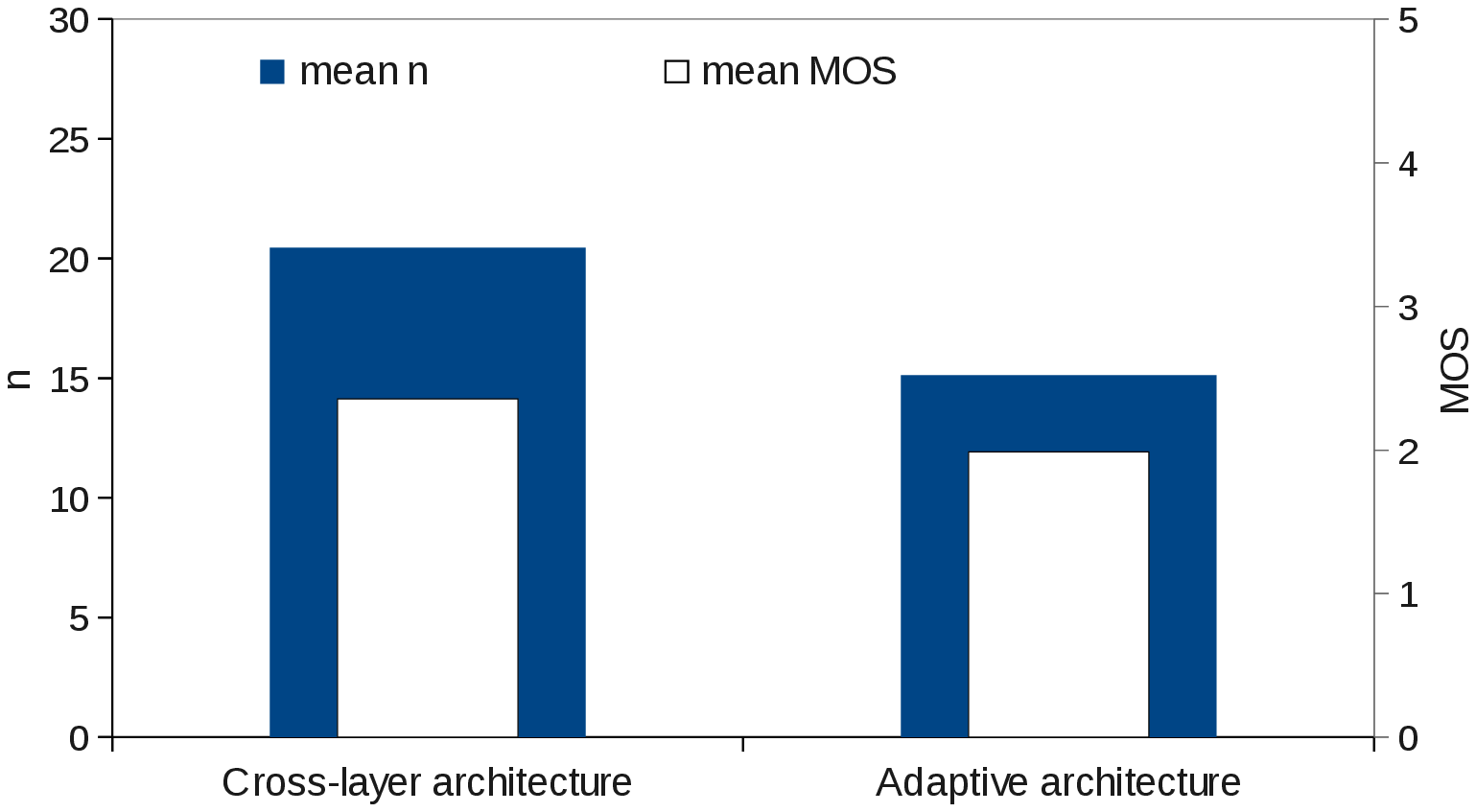}
\label{fig:mos-Session_crossANDadap_Grandma}}
\caption{Mean MOS of the video flows and mean \emph{number of sessions} in the \emph{cross-layer architecture} and \emph{adaptive architecture} for MAD and Grandma sequences}
\label{fig:mos-Session_crossANDadap}
\end{figure}

To compare the difference between the mean MOS of the video flows and mean \emph{number of sessions} in the \emph{cross-layer architecture} and \emph{adaptive architecture}, both are plotted in the bar charts in Figure  \ref{fig:mos-Session_crossANDadap}. The white bars represent the mean MOS and blue bars represent the mean \emph{number of sessions}.

\begin{figure}[t]
\centering
\subfigure[MAD (CIF)]{%
\includegraphics[height=4cm,width=5.7cm]{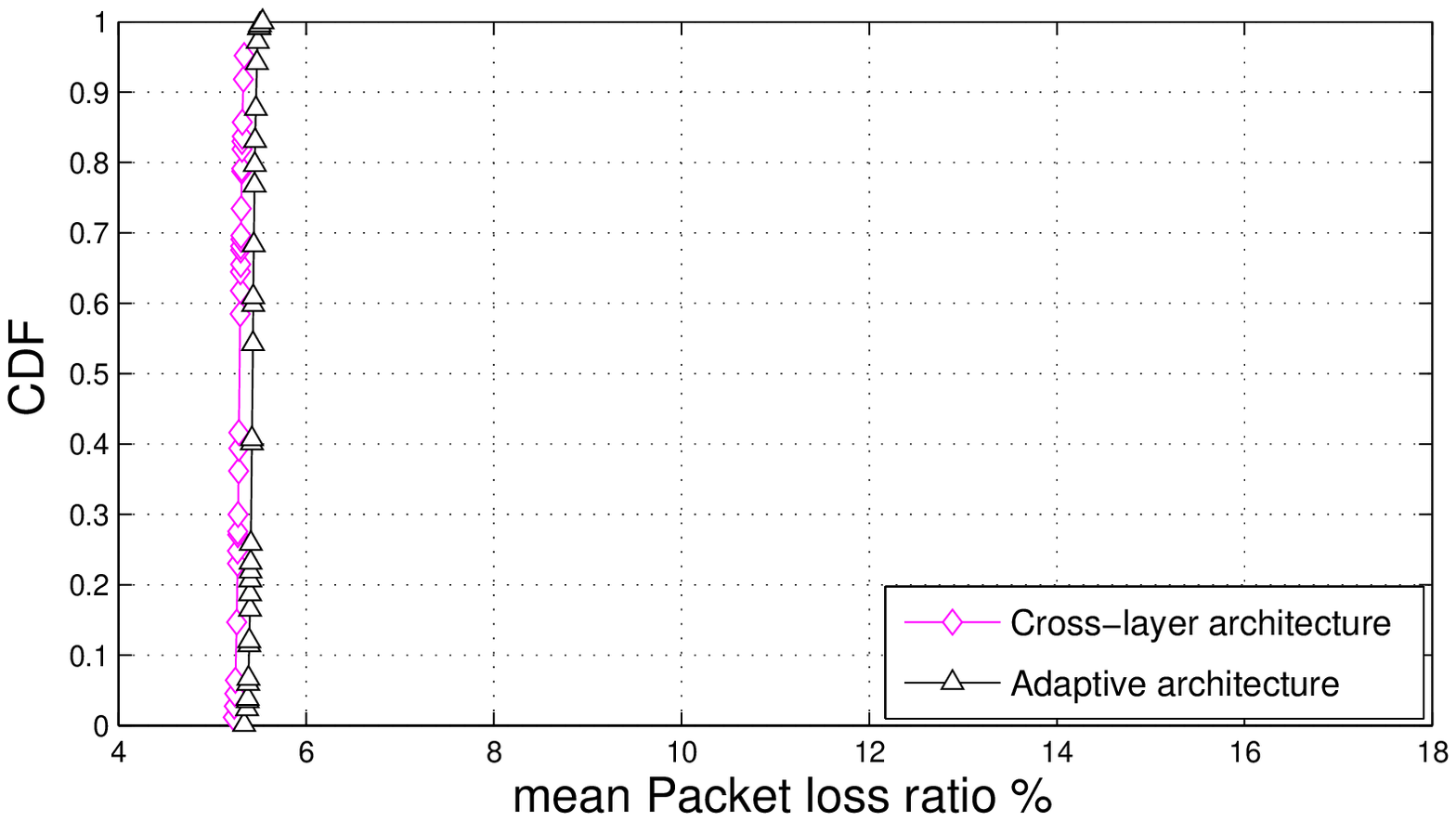}
\label{fig:meanPktLossRatio_crossANDadap_MAD}}
\quad
\subfigure[Grandma (QCIF)]{%
\includegraphics[height=4cm,width=5.7cm]{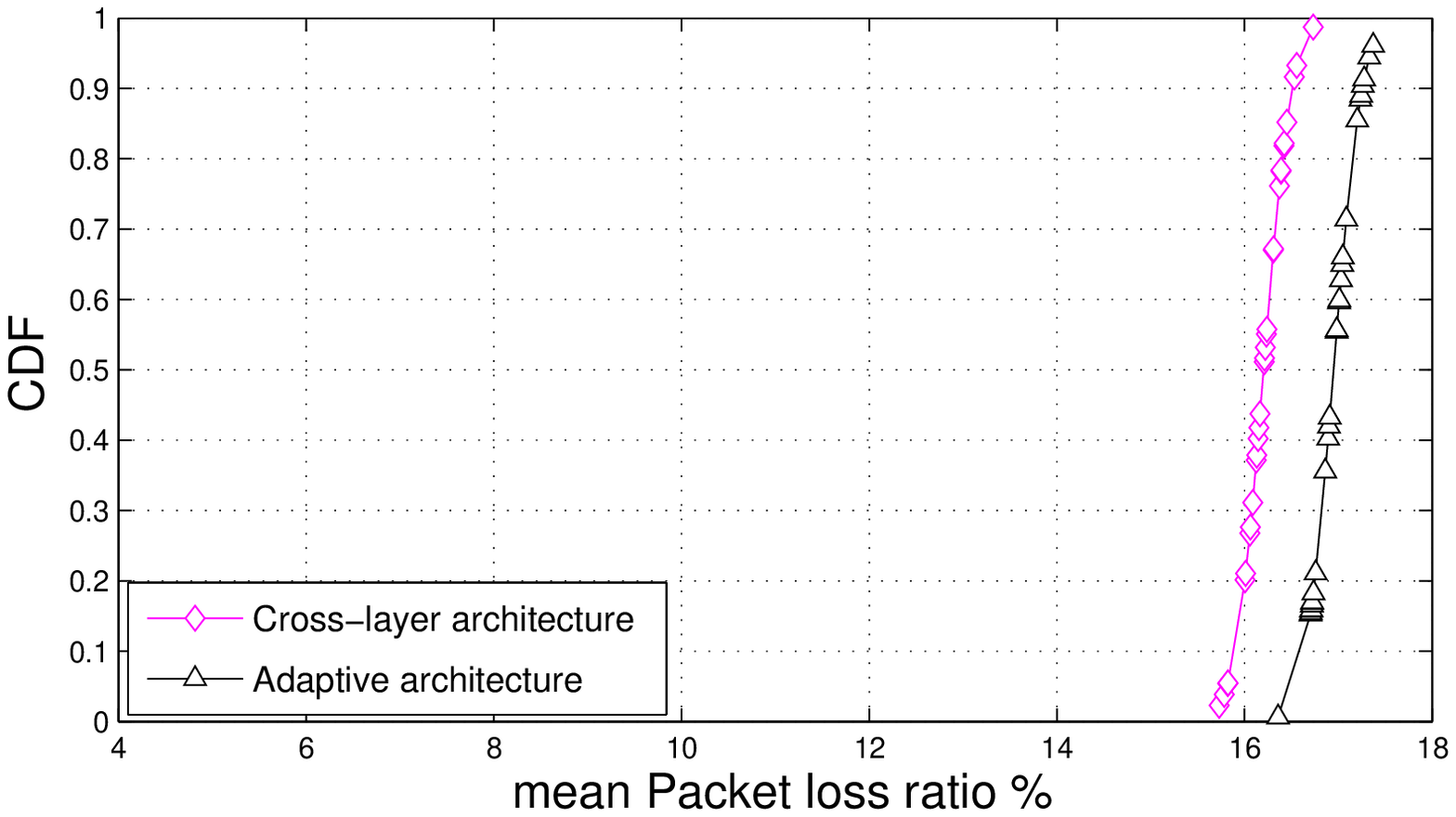}
\label{fig:meanPktLossRatio_crossANDadap_Grandma}}
\caption{CDF of the mean packet loss ratio of the video flows in the \emph{cross-layer architecture} and \emph{adaptive architecture} for MAD and Grandma sequences}
\label{fig:meanPktLossRatio_crossANDadap}
\end{figure}

\begin{figure}[t]
\centering
\subfigure[MAD (CIF)]{%
\includegraphics[height=4cm,width=5.7cm]{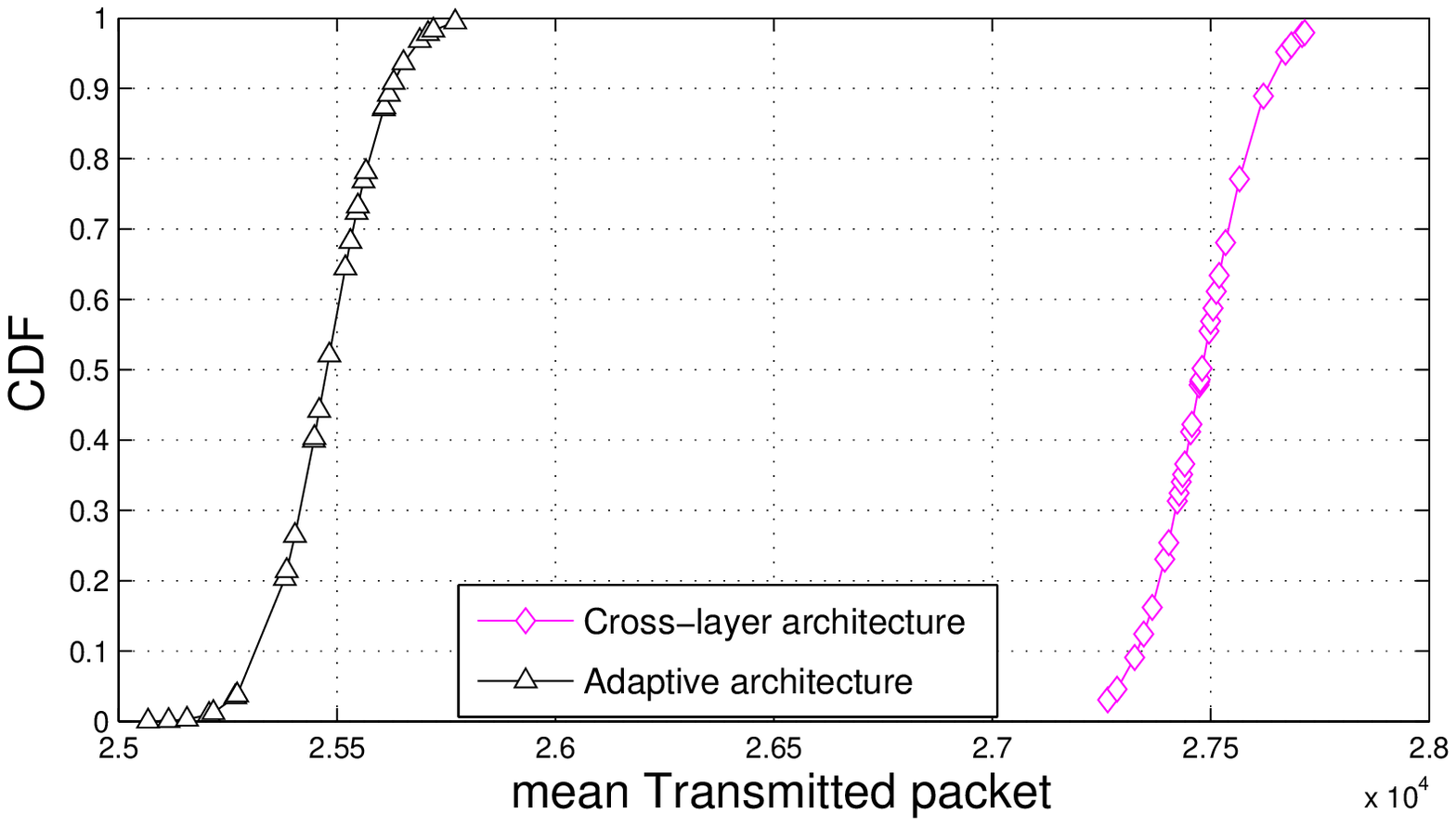}
\label{fig:meanTransmittedPkt_crossANDadap_MAD}}
\quad
\subfigure[Grandma (QCIF)]{%
\includegraphics[height=4cm,width=5.7cm]{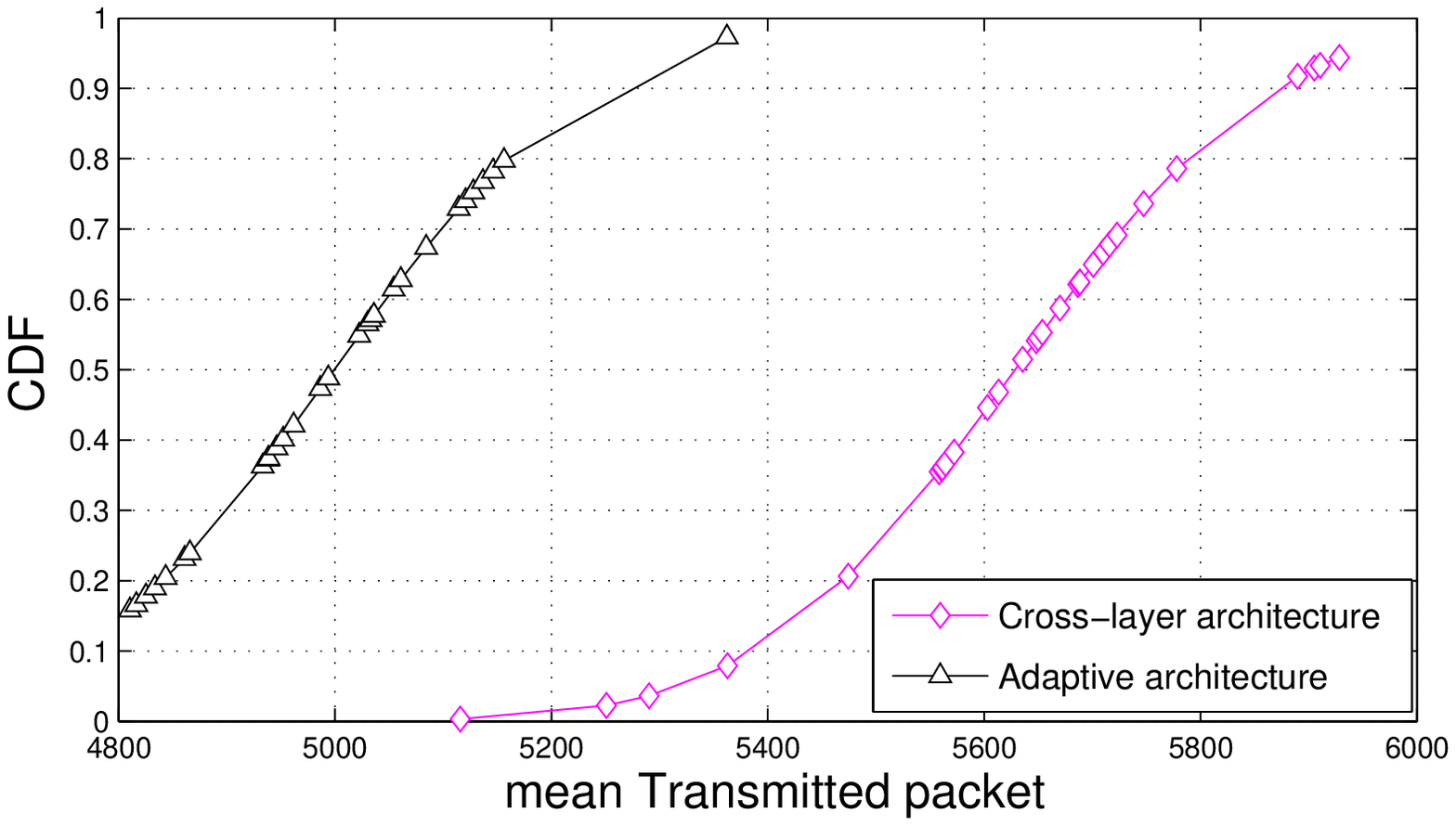}
\label{fig:meanTransmittedPkt_crossANDadap_Grandma}}
\caption{CDF of the mean transmitted packet of the video flows in the \emph{cross-layer architecture} and \emph{adaptive architecture} for MAD and Grandma sequences}
\label{fig:meanTransmittedPkt_crossANDadap}
\end{figure}

Video streaming services are tolerant to packet loss to some extent. Error concealment in the decoder allows video to accept some tolerance of packet loss. We calculated the CDF of the mean packet drop ratio of the video flows in the \emph{cross-layer architecture} and \emph{adaptive architecture} and plotted them in Figure \ref{fig:meanPktLossRatio_crossANDadap}. The video flows delivered over the \emph{cross-layer architecture} experienced less packet drop compared to the video flows in the \emph{adaptive architecture}.

\begin{figure}[t]
\centering
\subfigure[MAD (CIF)]{%
\includegraphics[height=4cm,width=5.7cm]{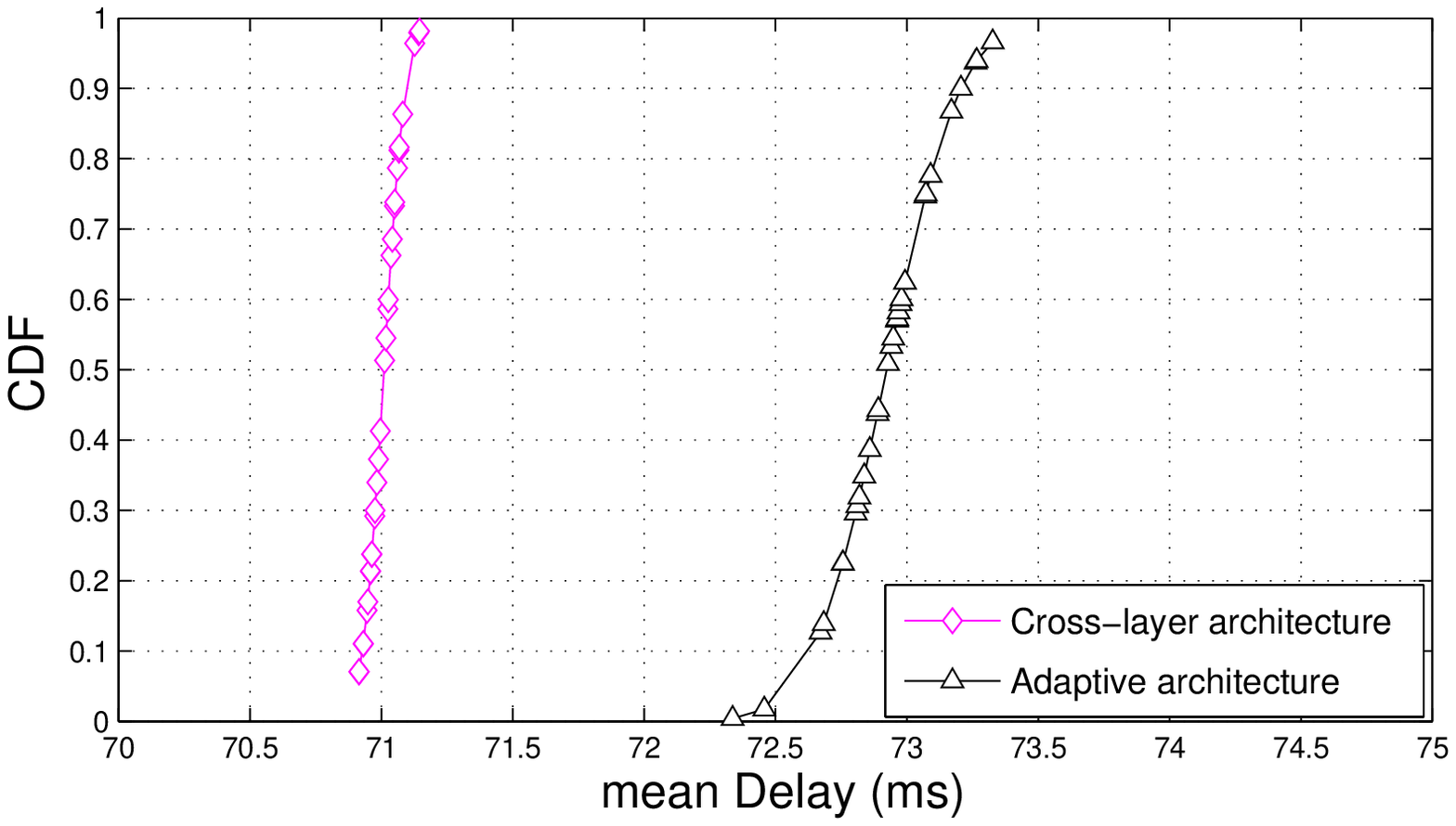}
\label{fig:meanDelay_adapANDnonadap_MAD}}
\quad
\subfigure[Grandma (QCIF)]{%
\includegraphics[height=4cm,width=5.7cm]{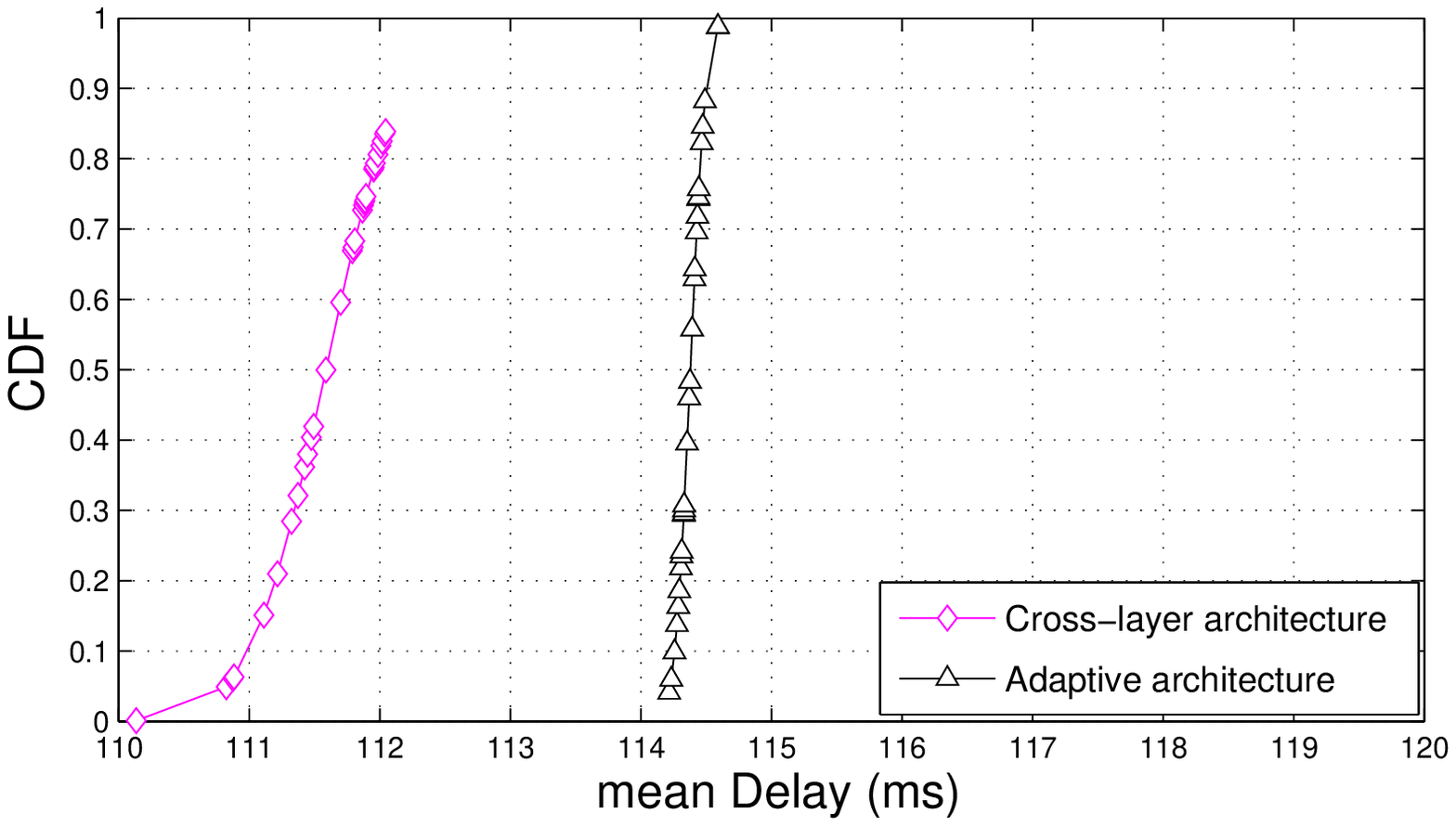}
\label{fig:meanDelay_adapANDnonadap_Grandma}}
\caption{CDF of the mean delay of the video flows in the \emph{cross-layer architecture} and \emph{adaptive architecture} for MAD and Grandma sequences}
\label{fig:meanDelay_crossANDadap}
\end{figure}

In contrast to the substantial difference in the mean MOS as shown in Figure \ref{fig:meanMOS_cross_adaptive}, there is a small difference between the packet drop ratio of the video flows in the \emph{cross-layer architecture} and \emph{adaptive architecture} as can be seen in Figure \ref{fig:meanPktLossRatio_crossANDadap}. However, these packets were dropped out of the total number of the transmitted packets. The CDF of the mean transmitted packet are shown in Figure \ref{fig:meanTransmittedPkt_crossANDadap} in which the difference between the number of packets transmitted by the video sources in each of the \emph{cross-layer architecture} and \emph{adaptive architecture} is evident. Therefore, a smaller ratio of the packet loss of the video flows out of a higher number of transmitted packets of the same video content in the \emph{cross-layer architecture} compared to the \emph{adaptive architecture} ensured a better quality (in terms of MOS) as discussed earlier in this section.

From Equation (\ref{eq:QoE_utlityFunction}), it is evident that a higher bitrate provides a better MOS for the same packet drop ratio. Sending a higher number of video packets by the \emph{cross-layer architecture} compared to \emph{adaptive architecture} as shown in Figure \ref{fig:meanTransmittedPkt_crossANDadap} and a lower packet drop ratio as shown in Figure \ref{fig:meanPktLossRatio_crossANDadap} over the same simulation time (500 seconds), indicates that the video content was sent with a higher bitrate, thus a better MOS was provided by the \emph{cross-layer architecture}.

\begin{figure}[t]
\centering
\subfigure[MAD (CIF)]{%
\includegraphics[height=4cm,width=5.7cm]{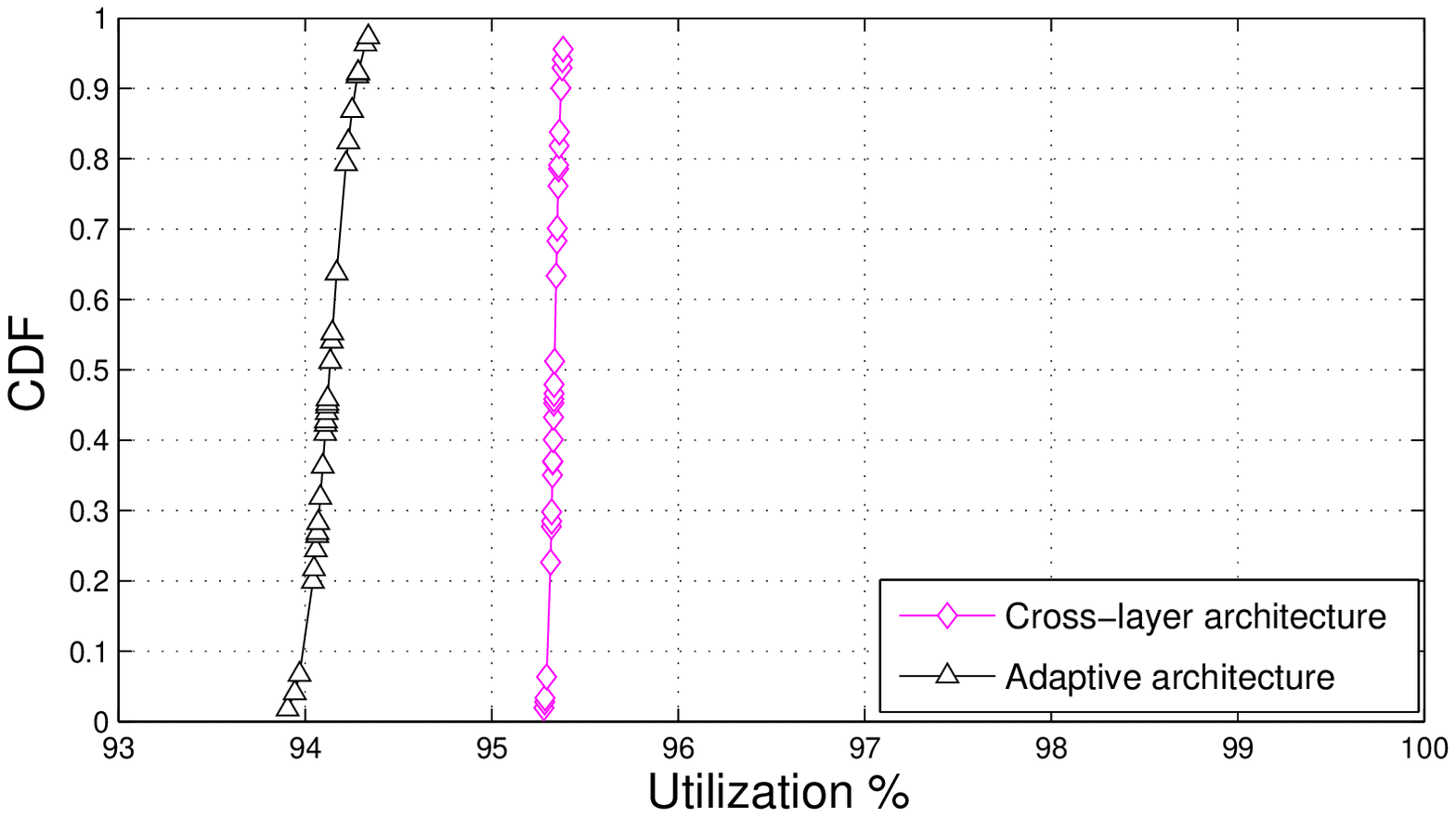}
\label{fig:utilisation_crossANDadap_MAD}}
\quad
\subfigure[Grandma (QCIF)]{%
\includegraphics[height=4cm,width=5.7cm]{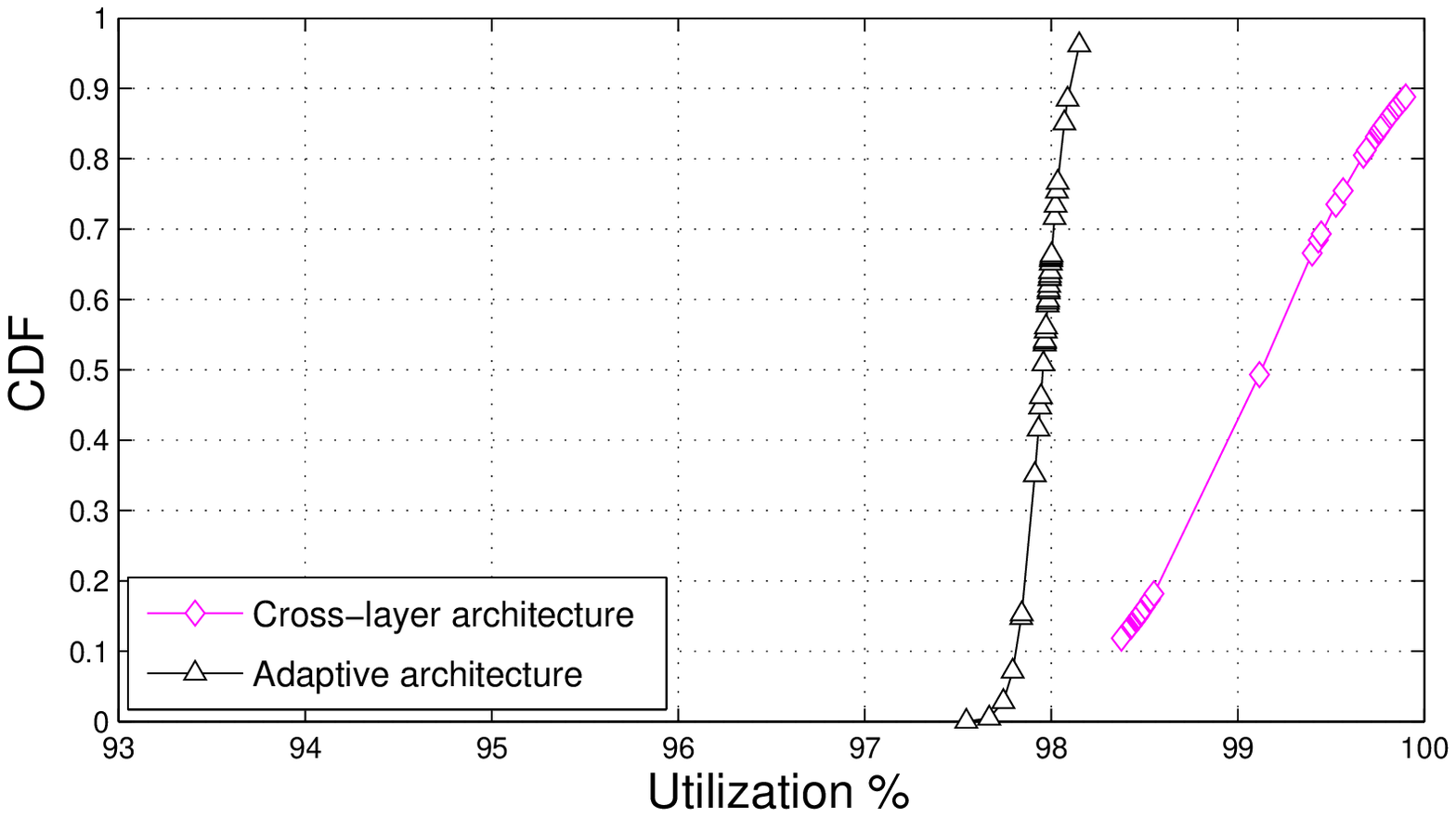}
\label{fig:utilisation_crossANDadap_Grandma}}
\caption{Utilization of the \emph{cross-layer architecture} and \emph{adaptive architecture} for MAD and Grandma sequences}
\label{fig:utilisation_crossANDadap}
\end{figure}

Video streaming applications have a lenient delay requirement. Depending on the application's buffering capabilities, 4 to 5 seconds delay is acceptable \cite{Szigeti2004}. The CDF of the mean delay of the video flows for each of the \emph{cross-layer architecture} and \emph{adaptive architecture} was measured. Figure \ref{fig:meanDelay_crossANDadap} show that the video flows in the \emph{cross-layer architecture} experienced less delay compared to the video flows in the \emph{adaptive architecture}. 

\begin{figure}[t]
\centering
\subfigure[MAD (CIF)]{%
\includegraphics[height=4cm,width=5.7cm]{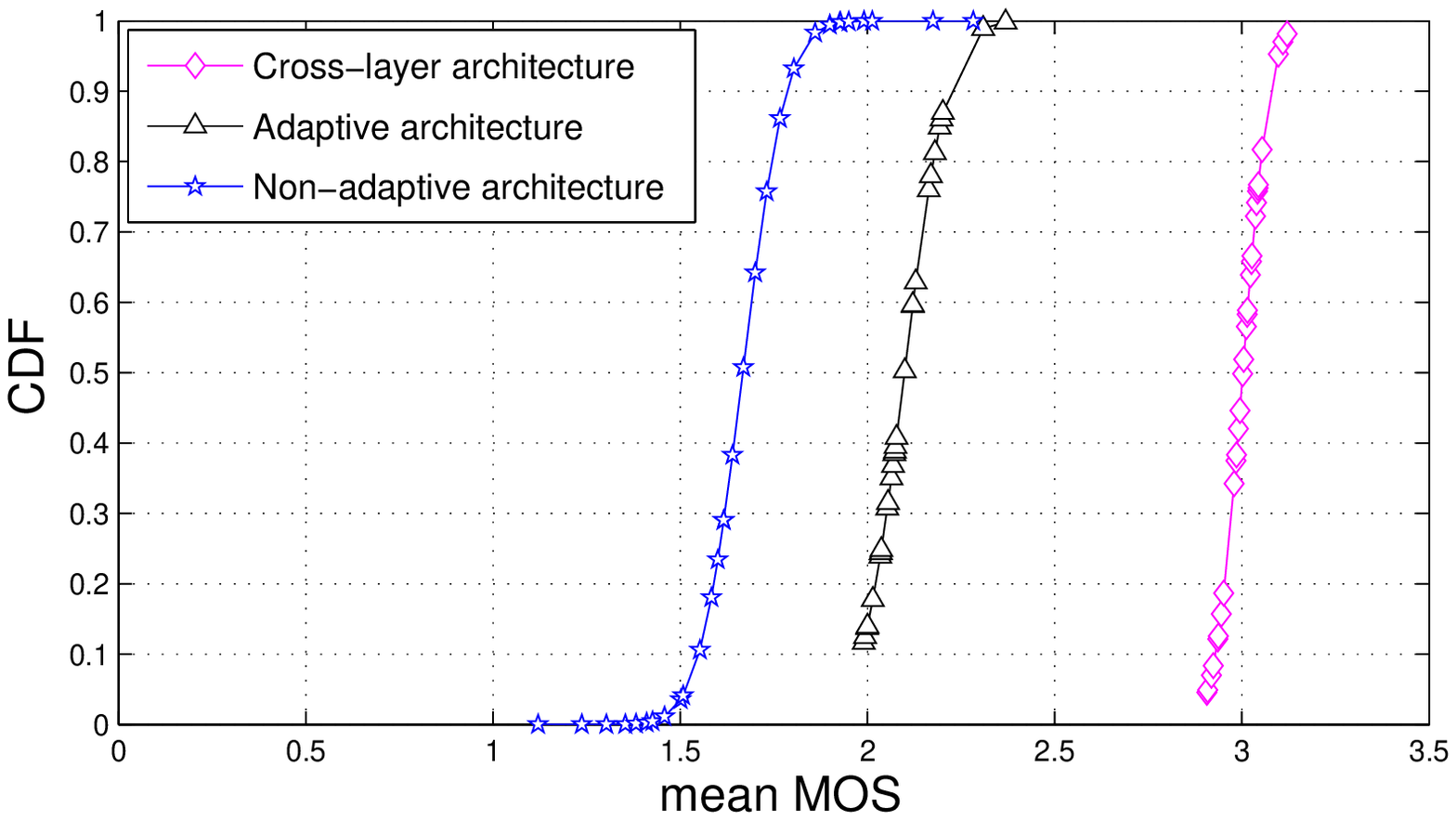}
\label{fig:meanMOS_all_MAD}}
\quad
\subfigure[Grandma (QCIF)]{%
\includegraphics[height=4cm,width=5.7cm]{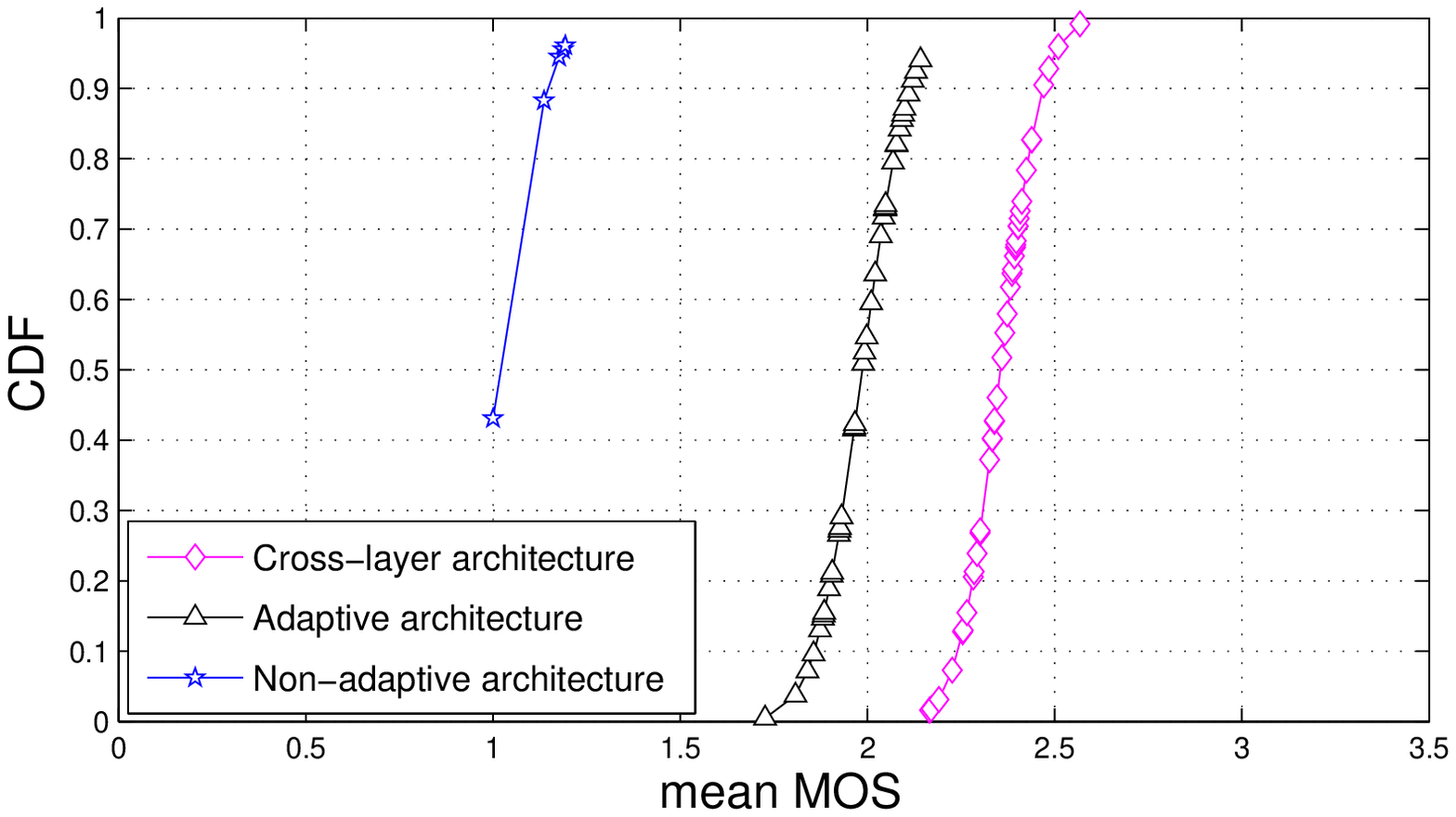}
\label{fig:meanMOS_all_Grandma}}
\caption{CDF of the mean MOS of the video flows in the \emph{cross-layer architecture}, \emph{adaptive architecture} and \emph{non-adaptive architecture} for MAD and Grandma sequences}
\label{fig:meanMOS_all}
\end{figure}

The \emph{adaptive architecture} utilizes the capacity of the bottleneck link less efficiently than the \emph{cross-layer architecture} as can be observed in Figure \ref{fig:utilisation_crossANDadap}. Note that the utilization measure includes the FTP flows also. It is calculated as the number of transmitted bits over the capacity of the link for the simulation period. Thus, the \emph{adaptive architecture} led to a high link utilization; 94\% for MAD sequence and 98\% for Grandma sequence. The utilization of the \emph{cross-layer architecture} however, increased to 95\% for MAD sequence and 99\% for Grandma sequence. We can conclude that the utilization figures can not decide the performance of the two architectures for the video flows as it is calculated for video and FTP flows.

Finally, the video flows delivered over the proposed \emph{cross-layer architecture} are compared to the video flows transmitted by each of the \emph{adaptive architecture} and \emph{non-adaptive architecture}. Figure \ref{fig:meanMOS_all} shows the CDF of the mean MOS of the video flows in the three architectures for both video resolutions. While, there is an improvement of the mean MOS of the video flows in the \emph{adaptive architecture} through the adaptation of the sender rate compared to the video flows in the \emph{non-adaptive architecture}, a higher value of the mean MOS of the video flows in the \emph{cross-layer architecture} is observed.

\begin{figure}[t]
\centering
\subfigure[MAD (CIF)]{%
\includegraphics[height=4cm,width=5.7cm]{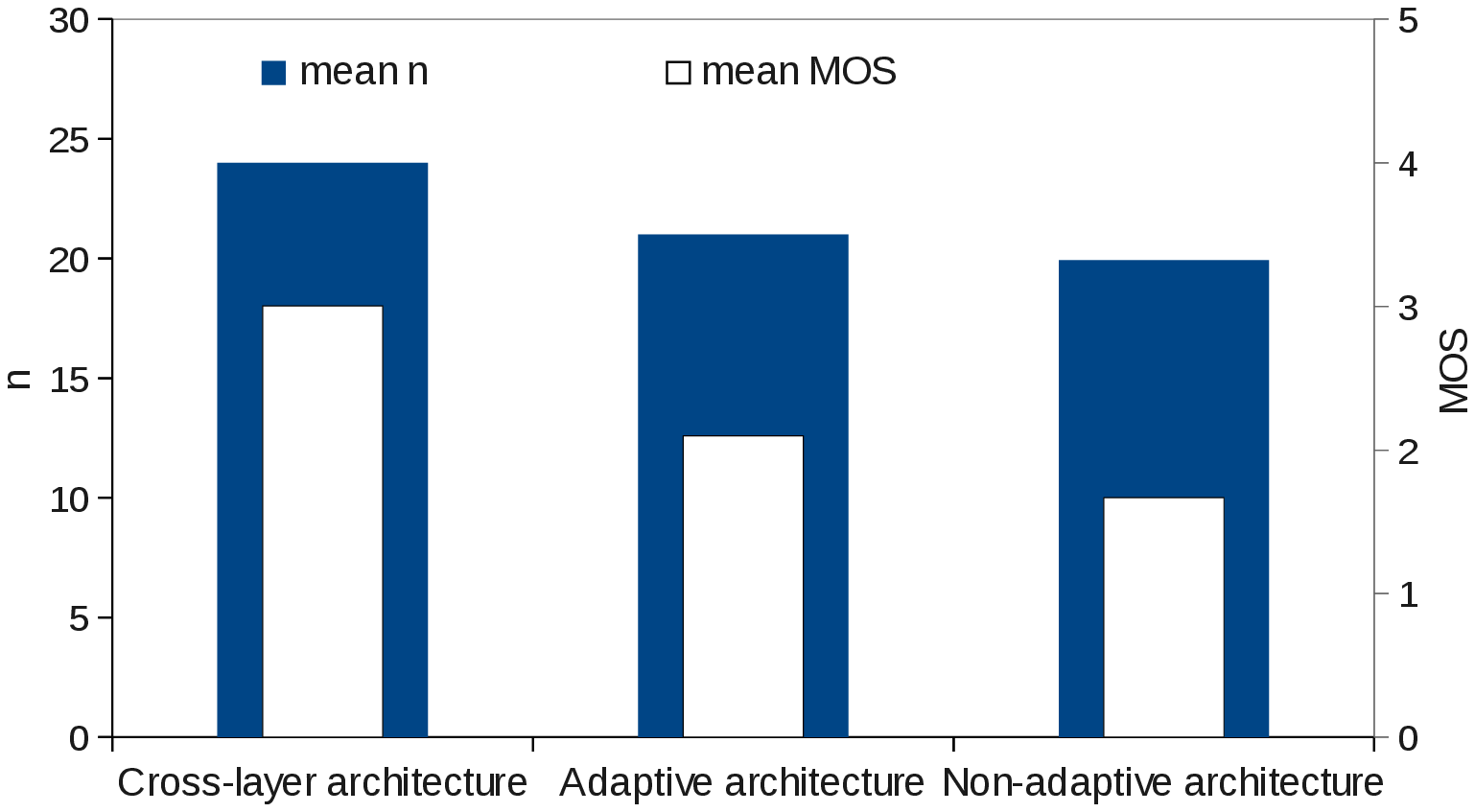}
\label{fig:mos-Session_all_MAD}}
\quad
\subfigure[Grandma (QCIF)]{%
\includegraphics[height=4cm,width=5.7cm]{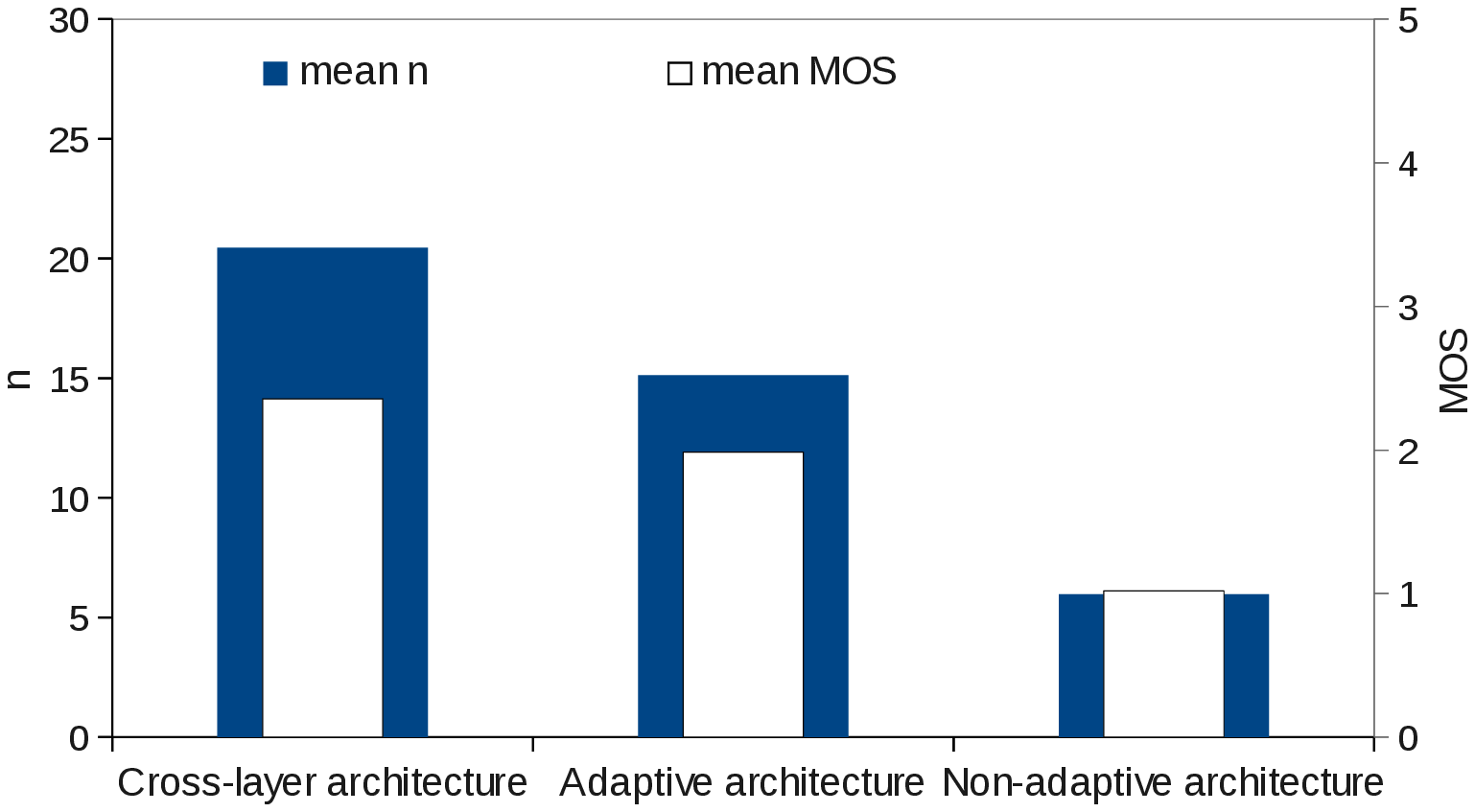}
\label{fig:mos-Session_all_Grandma}}
\caption{Mean MOS of the video flows and mean \emph{number of sessions} in the \emph{cross-layer architecture}, \emph{adaptive architecture} and \emph{non-adaptive architecture} for MAD and Grandma sequences}
\label{fig:mos-Session_all}
\end{figure}

Moreover, the proposed \emph{cross-layer architecture} accepts and delivers a higher \emph{number of sessions} compared to the other two architectures (\emph{adaptive architecture} and \emph{non-adaptive architecture}). This can be observed in Figure \ref{fig:meanSession_all}. The bar chart in Figure \ref{fig:mos-Session_all} illustrates the difference in the mean MOS of the video flows and mean \emph{number of sessions} between all three architectures for both resolutions.

\section{Conclusion}
\label{Sec:Conclusion}
A QoE-aware \emph{cross-layer architecture} for video streaming services was proposed in this paper. A combination of the rate adaptation and QoE-aware admission control are two main components of the architecture. The performance of the \emph{cross-layer architecture} was analyzed and compared to two other architectures (\emph{adaptive architecture} and \emph{non-adaptive architecture}). The extensive simulation results have shown that the \emph{cross-layer architecture} can provide an improvement in the mean MOS, considerably higher number of successful decoded video session, less mean delay and packet loss. At the same time it utilizes the link more efficiently. Evaluating the architecture with a greater variety of video contents and developing Algorithm \ref{pro-architecture} to include post-acceptance bit rate switching are interesting areas of future research. Future studies may also consider higher resolutions for the evaluation purpose. Another interesting area of research is to have a dynamic number of video variants instead of transcoding each content into a fixed number of video files.

\section*{References}
\bibliographystyle{elsarticle-num}

\begin{thebibliography}{10}
\expandafter\ifx\csname url\endcsname\relax
  \def\url#1{\texttt{#1}}\fi
\expandafter\ifx\csname urlprefix\endcsname\relax\def\urlprefix{URL }\fi
\expandafter\ifx\csname href\endcsname\relax
  \def\href#1#2{#2} \def\path#1{#1}\fi

\bibitem{Cisco2014a}
{Cisco documentation}, Cisco visual networking index: Forecast and methodology
  2013-2018, {Cisco white paper} (Jun. 2014).

\bibitem{ITU-T2007}
{ITU-T Document FG IPTV-IL-0050}, {Definition of quality of experience (QoE)}
  (Jan. 2007).

\bibitem{Brooks2010}
P.~Brooks, B.~Hestnes, User measures of quality of experience: Why being
  objective and quantitative is important, Network, IEEE 24~(2) (2010) 8 --13.

\bibitem{Fu2013}
B.~Fu, D.~Munaretto, T.~Melia, B.~Sayadi, W.~Kellerer, Analyzing the
  combination of different approaches for video transport optimization for next
  generation cellular networks, Network, IEEE 27~(2) (2013) 8--14.

\bibitem{Qadir2013}
S.~Qadir, A.~Kist, Quality of experience enhancement through adapting sender
  bit rate, in: TENCON Spring Conference, 2013 IEEE, 2013, pp. 490--494.

\bibitem{Qadir2015}
Q.~M. Qadir, A.~A. Kist, Z.~Zhang, A novel traffic rate measurement algorithm
  for quality of experience-aware video admission control, Multimedia, IEEE
  Transactions on 17~(5) (2015) 711--722.

\bibitem{Qadir2015a}
Q.~M. Qadir, A.~A. Kist, Z.~Zhang, Mechanisms for {QoE} optimisation of video
  traffic: a review paper, Australasian Journal of Information, Communication
  Technology and Applications 1~(2) (2015) 1--18.

\bibitem{Ernst2014}
J.~B. Ernst, S.~C. Kremer, J.~J. Rodrigues, A survey of {QoS/QoE} mechanisms in
  heterogeneous wireless networks, Physical Communication 13, Part B~(0) (2014)
  61--72, special Issue on Heterogeneous and Small Cell Networks.

\bibitem{Maallawi2014}
R.~Maallawi, N.~Agoulmine, B.~Radier, T.~ben Meriem, A comprehensive survey on
  offload techniques and management in wireless access and core networks,
  Communications Surveys Tutorials, IEEE PP~(99) (2014) 1--1.

\bibitem{Goudarzi2012}
P.~Goudarzi, Scalable video transmission over multi-hop wireless networks with
  enhanced quality of experience using {Swarm} intelligence, Signal Processing:
  Image Communication 27~(7) (2012) 722--736.

\bibitem{Goudarzi2010}
P.~Goudarzi, M.~Hosseinpour, Video transmission over {MANETs} with enhanced
  quality of experience, Consumer Electronics, IEEE Transactions on 56~(4)
  (2010) 2217--2225.

\bibitem{Politis2012}
I.~Politis, L.~Dounis, T.~Dagiuklas, {H.264/SVC} vs. {H.264/AVC} video quality
  comparison under {QoE}-driven seamless handoff, Signal Processing: Image
  Communication 27~(8) (2012) 814--826.

\bibitem{Khan2006}
S.~Khan, Y.~Peng, E.~Steinbach, M.~Sgroi, W.~Kellerer, Application-driven
  cross-layer optimization for video streaming over wireless networks,
  Communications Magazine, IEEE 44~(1) (2006) 122--130.

\bibitem{Thakolsri2009}
S.~Thakolsri, S.~Khan, E.~Steinbach, W.~Kellerer, {QoE}-driven cross-layer
  optimization for high speed downlink packet access, Journal of Communications
  4~(9) (2009) 669--680.

\bibitem{Khalek2012}
A.~Khalek, C.~Caramanis, R.~Heath, A cross-layer design for perceptual
  optimization of {H.264/SVC} with unequal error protection, Selected Areas in
  Communications, IEEE Journal on 30~(7) (2012) 1157--1171.

\bibitem{Chen2014}
X.~Chen, J.-N. Hwang, C.-N. Lee, S.-I. Chen, A near optimal {QoE}-driven power
  allocation scheme for scalable video transmissions over {MIMO} systems,
  Selected Topics in Signal Processing, IEEE Journal of PP~(99) (2014) 1--1.

\bibitem{Latre2011}
S.~Latr{\'e}, Autonomic {QoE} management of multimedia networks, Phd
  dissertation, Universiteit Gent (Jun. 2011).

\bibitem{Latre2013}
S.~Latr{\'e}, F.~De~Turck, Joint in-network video rate adaptation and
  measurement-based admission control: Algorithm design and evaluation, Journal
  of Network and Systems Management 21~(4) (2013) 588--622.

\bibitem{Chen2015}
C.~Chen, X.~Zhu, G.~de~Veciana, A.~Bovik, R.~Heath, Rate adaptation and
  admission control for video transmission with subjective quality constraints,
  Selected Topics in Signal Processing, IEEE Journal of 9~(1) (2015) 22--36.

\bibitem{Debono2012}
C.~Debono, B.~Micallef, N.~Philip, A.~Alinejad, R.~Istepanian, N.~Amso,
  Cross-layer design for optimized region of interest of ultrasound video data
  over mobile {WiMAX}, Information Technology in Biomedicine, IEEE Transactions
  on 16~(6) (2012) 1007--1014.

\bibitem{Singhal2014}
C.~Singhal, S.~De, R.~Trestian, G.-M. Muntean, Joint optimization of
  user-experience and energy-efficiency in wireless multimedia broadcast,
  Mobile Computing, IEEE Transactions on 13~(7) (2014) 1522--1535.

\bibitem{Rugelj2014}
M.~Rugelj, U.~Sedlar, M.~Volk, J.~Sterle, M.~Hajdinjak, A.~Kos, Novel
  cross-layer {QoE}-aware radio resource allocation algorithms in multiuser
  {OFDMA} systems, Communications, IEEE Transactions on 62~(9) (2014)
  3196--3208.

\bibitem{Ju2012}
Y.~Ju, Z.~Lu, W.~Zheng, X.~Wen, D.~Ling, A cross-layer design for video
  applications based on {QoE} prediction, in: Wireless Personal Multimedia
  Communications (WPMC), 2012 15th International Symposium on, 2012, pp.
  534--538.

\bibitem{Ivesic2014}
K.~Ivesic, L.~Skorin-Kapov, M.~Matijasevic, Cross-layer {QoE}-driven admission
  control and resource allocation for adaptive multimedia services in {LTE},
  Journal of Network and Computer Applications 46~(0) (2014) 336--351.

\bibitem{Zhou2013}
L.~Zhou, Z.~Yang, Y.~Wen, H.~Wang, M.~Guizani, Resource allocation with
  incomplete information for {QoE}-driven multimedia communications, Wireless
  Communications, IEEE Transactions on 12~(8) (2013) 3733--3745.

\bibitem{Khan2010}
A.~Khan, L.~Sun, E.~Jammeh, E.~Ifeachor, Quality of experience-driven
  adaptation scheme for video applications over wireless networks,
  Communications, IET 4~(11) (2010) 1337--1347.

\bibitem{Zhao2014}
M.~Zhao, X.~Gong, J.~Liang, W.~Wang, X.~Que, S.~Cheng, {QoE}-driven cross-layer
  optimization for wireless dynamic adaptive streaming of scalable videos over
  {HTTP}, Circuits and Systems for Video Technology, IEEE Transactions on
  PP~(99) (2014) 1--1.

\bibitem{ElEssaili2014}
A.~El~Essaili, D.~Schroeder, E.~Steinbach, D.~Staehle, M.~Shehada, {QoE}-based
  traffic and resource management for adaptive {HTTP} video delivery in {LTE},
  Circuits and Systems for Video Technology, IEEE Transactions on PP~(99)
  (2014) 1--1.

\bibitem{Fiedler2009}
M.~Fiedler, H.-J. Zepernick, L.~Lundberg, P.~Arlos, M.~Pettersson, {QoE}-based
  cross-layer design of mobile video systems: Challenges and concepts, in:
  Computing and Communication Technologies, 2009. RIVF '09. International
  Conference on, 2009, pp. 1--4.

\bibitem{Latre2009}
S.~Latr{\'e}, P.~Simoens, B.~De~Vleeschauwer, W.~Meerssche, F.~Turck,
  B.~Dhoedt, P.~Demeester, S.~Berghe, E.~Lumley, An autonomic architecture for
  optimizing {QoE} in multimedia access networks, Computer Networks 53~(10)
  (2009) 1587--1602.

\bibitem{Oyman2012}
O.~Oyman, S.~Singh, {Q}uality of experience for {HTTP} adaptive streaming
  services, Communications Magazine, IEEE 50~(4) (2012) 20--27.

\bibitem{Zhang2011}
J.~Zhang, N.~Ansari, On assuring end-to-end {QoE} in next generation networks:
  challenges and a possible solution, Communications Magazine, IEEE 49~(7)
  (2011) 185--191.

\bibitem{Mathieu2011}
B.~Mathieu, S.~Ellouze, N.~Schwan, D.~Griffin, E.~Mykoniati, T.~Ahmed,
  O.~Prats, Improving end-to-end {QoE} via close cooperation between
  applications and {ISPs}, Communications Magazine, IEEE 49~(3) (2011)
  136--143.

\bibitem{Lie2008}
A.~Lie, J.~Klaue, Evalvid-{RA}: trace driven simulation of rate adaptive
  {MPEG-4 VBR} video, Multimedia Systems 14 (2008) 33--50.

\bibitem{Latre2011b}
S.~Latr{\'e}, R.~Klaas, T.~Wauters, F.~DeTurck, Protecting video service
  quality in multimedia access networks through {PCN}, Communications Magazine,
  IEEE 49~(12) (2011) 94--101.

\bibitem{Hoeffding1963}
W.~Hoeffding, Probability inequalities for sums of bounded random variables,
  Journal of the American Statistical Association 58~(301) (1963) 13--30.

\bibitem{Floyd2008}
S.~Floyd, M.~Handley, J.~Padhye, J.~Widmer, {TCP} friendly rate control
  ({TFRC}){:} protocol specification, RFC 5348, IETF, [Oline]
  \url{https://www.ietf.org/rfc/rfc3448.txt}, accessed on: Nov. 7th, 2015 (Sep.
  2008).

\bibitem{Hamdi1997}
H.~Hamdi, J.~Roberts, P.~Rolin, Rate control for {VBR} video coders in
  broad-band networks, Selected Areas in Communications, IEEE Journal on 15~(6)
  (1997) 1040--1051.

\bibitem{Rosenberg2002}
J.~Rosenberg, H.~Schulzrinne, G.~Camarillo, A.~Johnston, J.~Peterson,
  R.~Sparks, M.~Handley, S.~E., {SIP: Session Initiation Protocol} (Jun. 2002).

\bibitem{Zhang2013b}
W.~Zhang, Y.~Wen, Z.~Chen, A.~Khisti, {QoE}-driven cache management for {HTTP}
  adaptive bit rate streaming over wireless networks, Multimedia, IEEE
  Transactions on 15~(6) (2013) 1431--1445.

\bibitem{ns2}
{NS-2}, {The Network Simulator}, [Online] http://www.isi.edu/nsnam/ns/,
  accessed on: May 28, 2015.

\bibitem{Khan2012}
A.~Khan, L.~Sun, E.~Ifeachor, {QoE} prediction model and its application in
  video quality adaptation over {UMTS} networks, Multimedia, IEEE Transactions
  on 14~(2) (2012) 431--442.

\bibitem{ffmpeg2004}
{FFMPEG M}ultimedia {S}ystem, [Online] \url{http://ffmpeg.mplayerhq.hu/},
  accessed on: Nov. 5th, 2015 (2004).

\bibitem{Chen2014a}
Y.~Chen, B.~Zhang, C.~Chen, D.~M. Chiu, Performance modeling and evaluation of
  peer-to-peer live streaming systems under flash crowds, Networking, IEEE/ACM
  Transactions on 22~(4) (2014) 1106--1120.

\bibitem{Szigeti2004}
T.~Szigeti, C.~Hattingh, {End-to-end QoS network design: Quality of service in
  LANs, WANs, and VPNs (Networking Technology)}, Cisco Press, Indianapolis, IN,
  USA, 2004.

\bibitem{Gross2004}
J.~Gross, J.~Klaue, H.~Karl, A.~Wolisz, Cross-layer optimization of {OFDM}
  transmission systems for {MPEG}-4 video streaming, Computer Communications
  27~(11) (2004) 1044--1055.

\bibitem{Papadimitriou2007}
P.~Papadimitriou, V.~Tsaoussidis, {SSVP}: A congestion control scheme for
  real-time video streaming, Computer Networks 51~(15) (2007) 4377--4395.

\bibitem{Li2010}
D.~Li, J.~Pan, Performance evaluation of video streaming over multi-hop
  wireless local area networks, Wireless Communications, IEEE Transactions on
  9~(1) (2010) 338--347.

\bibitem{Zheng2015}
K.~Zheng, X.~Zhang, Q.~Zheng, W.~Xiang, L.~Hanzo, Quality-of-experience
  assessment and its application to video services in {LTE} networks, Wireless
  Communications, IEEE 22~(1) (2015) 70--78.

\bibitem{Khan2010a}
A.~Khan, L.~Sun, E.~Ifeachor, Learning models for video quality prediction over
  wireless local area network and universal mobile telecommunication system
  networks, Communications, IET 4~(12) (2010) 1389--1403.

\bibitem{Ohm2004}
J.-R. Ohm, Multimedia communication technology, Vol.~1, Springer, 2004.

\bibitem{Stankiewicz2011}
R.~Stankiewicz, P.~Cholda, A.~Jajszczyk, {QoX}: what is it really?,
  Communications Magazine, IEEE 49~(4) (2011) 148--158.

\end{thebibliography}

\end{document}